\documentclass[usegraphicx,useAMS,usenatbib]{mn2e}
\usepackage{times}
\usepackage{amssymb}
\usepackage{amsmath}
\usepackage{graphicx}
\usepackage{euscript}
\usepackage{rotating}
\usepackage{epsfig}

\def\ghz{{\rm\thinspace GHz}}

\def\kev{{\rm\thinspace keV}}

\def\arcsec{{\rm\thinspace arcsec}}

\begin{document}

\newcommand{\Mpc}{\rm\thinspace Mpc}
\newcommand{\kpc}{\rm\thinspace kpc}
\newcommand{\pc}{\rm\thinspace pc}
\newcommand{\km}{\rm\thinspace km}
\newcommand{\m}{\rm\thinspace m}
\newcommand{\cm}{\rm\thinspace cm}
\newcommand{\cmps}{\hbox{$\cm\s^{-1}\,$}}
\newcommand{\cmpssq}{\hbox{$\cm\s^{-2}\,$}}
\newcommand{\cmsq}{\hbox{$\cm^2\,$}}
\newcommand{\cmcu}{\hbox{$\cm^3\,$}}
\newcommand{\pcmcu}{\hbox{$\cm^{-3}\,$}}
\newcommand{\pcmcuK}{\hbox{$\cm^{-3}\K\,$}}

\newcommand{\yr}{\rm\thinspace yr}
\newcommand{\gyr}{\rm\thinspace Gyr}
\newcommand{\s}{\rm\thinspace s}
\newcommand{\ks}{\rm\thinspace ks}

\newcommand{\GHz}{\rm\thinspace GHz}
\newcommand{\MHz}{\rm\thinspace MHz}
\newcommand{\Hz}{\rm\thinspace Hz}

\newcommand{\K}{\rm\thinspace K}

\newcommand{\Kpcmc}{\hbox{$\K\cm^{-3}\,$}}

\newcommand{\g}{\rm\thinspace g}
\newcommand{\gpcm}{\hbox{$\g\cm^{-3}\,$}}
\newcommand{\gpcmps}{\hbox{$\g\cm^{-3}\s^{-1}\,$}}
\newcommand{\gps}{\hbox{$\g\s^{-1}\,$}}
\newcommand{\Msun}{\hbox{$\rm\thinspace M_{\odot}$}}
\newcommand{\Msunpc}{\hbox{$\Msun\pc^{-3}\,$}}
\newcommand{\Msunpkpc}{\hbox{$\Msun\kpc^{-1}\,$}}
\newcommand{\Msunppc}{\hbox{$\Msun\pc^{-3}\,$}}
\newcommand{\Msunppcpyr}{\hbox{$\Msun\pc^{-3}\yr^{-1}\,$}}
\newcommand{\Msunpyr}{\hbox{$\Msun\yr^{-1}\,$}}

\newcommand{\MeV}{\rm\thinspace MeV}
\newcommand{\keV}{\rm\thinspace keV}
\newcommand{\eV}{\rm\thinspace eV}
\newcommand{\erg}{\rm\thinspace erg}
\newcommand{\Jy}{\rm Jy}
\newcommand{\ergpcmc}{\hbox{$\erg\cm^{-3}\,$}}
\newcommand{\ergcmcups}{\hbox{$\erg\cm^3\ps\,$}}
\newcommand{\ergpcmps}{\hbox{$\erg\cm^{-3}\s^{-1}\,$}}
\newcommand{\ergpcmsqps}{\hbox{$\erg\cm^{-2}\s^{-1}\,$}}
\newcommand{\ergpcmsqpspA}{\hbox{$\erg\cm^{-2}\s^{-1}$\AA$^{-1}\,$}}
\newcommand{\ergpcmsqpspsr}{\hbox{$\erg\cm^{-2}\s^{-1}\sr^{-1}\,$}}
\newcommand{\ergpcmcups}{\hbox{$\erg\cm^{-3}\s^{-1}\,$}}
\newcommand{\ergps}{\hbox{$\erg\s^{-1}\,$}}
\newcommand{\ergpspmp}{\hbox{$\erg\s^{-1}\Mpc^{-3}\,$}}
\newcommand{\keVpcmsqpspsr}{\hbox{$\keV\cm^{-2}\s^{-1}\sr^{-1}\,$}}

\newcommand{\dyn}{\rm\thinspace dyn}
\newcommand{\dynpcmsq}{\hbox{$\dyn\cm^{-2}\,$}}

\newcommand{\kmps}{\hbox{$\km\s^{-1}\,$}}
\newcommand{\kmpspmp}{\hbox{$\km\s^{-1}\Mpc{-1}\,$}}
\newcommand{\kmpspMpc}{\hbox{$\kmps\Mpc^{-1}$}}

\newcommand{\Lsun}{\hbox{$\rm\thinspace L_{\odot}$}}
\newcommand{\Lsunppc}{\hbox{$\Lsun\pc^{-3}\,$}}

\newcommand{\Zsun}{\hbox{$\rm\thinspace Z_{\odot}$}}
\newcommand{\gauss}{\rm\thinspace gauss}
\newcommand{\arcsecond}{\rm\thinspace arcsec}
\newcommand{\chisq}{\hbox{$\chi^2$}}
\newcommand{\delchi}{\hbox{$\Delta\chi$}}
\newcommand{\ph}{\rm\thinspace ph}
\newcommand{\sr}{\rm\thinspace sr}

\newcommand{\pccm}{\hbox{$\cm^{-3}\,$}}
\newcommand{\psqcm}{\hbox{$\cm^{-2}\,$}}
\newcommand{\pcmsq}{\hbox{$\cm^{-2}\,$}}
\newcommand{\pmpc}{\hbox{$\Mpc^{-1}\,$}}
\newcommand{\pmpccu}{\hbox{$\Mpc^{-3}\,$}}
\newcommand{\ps}{\hbox{$\s^{-1}\,$}}
\newcommand{\pHz}{\hbox{$\Hz^{-1}\,$}}
\newcommand{\pcmK}{\hbox{$\cm^{-3}\K$}}
\newcommand{\phpcmsqps}{\hbox{$\ph\cm^{-2}\s^{-1}\,$}}
\newcommand{\psr}{\hbox{$\sr^{-1}\,$}}
\newcommand{\pspsqas}{\hbox{$\s^{-1}\,\arcsecond^{-2}\,$}}

\newcommand{\ergpspcmpK}{\hbox{$\erg\s^{-1}\cm^{-1}\K^{-1}\,$}}

\title{Investigating AGN Heating in a Sample of Nearby Clusters}
\author[Dunn \& Fabian]
{\parbox[]{6.in} {R.J.H. Dunn\thanks{E-mail: rjhd2@ast.cam.ac.uk} and A.C. Fabian\\
    \footnotesize
    Institute of Astronomy, Madingley Road, Cambridge CB3 0HA\\
  }}

\voffset-1cm

\maketitle

\begin{abstract}
We analyse those objects in the Brightest 55 sample of clusters of
galaxies which have a
short central cooling time and a central temperature drop.  Such
clusters are likely to require some form of heating.  Where clear
radio bubbles are observed in these clusters, their energy injection is compared
to the X-ray cooling rate.  Of the 20 clusters requiring heating, at
least 14 have clear bubbles, implying a duty cycle for the bubbling
activity of at least 70 per
cent.  The average distance out to which the bubbles can
offset the X-ray cooling, $r_{\rm heat}$, is given by $r_{\rm heat}/r_{\rm cool}=0.86\pm0.11$
where $r_{\rm cool}$ is defined as the radius as which the radiative
cooling time is $3\gyr$. 10 out of 16 clusters have
$r_{\rm heat}/r_{\rm cool}\gtrsim1$, but there is a large range in values.  The clusters which
require heating but show {\it no} clear bubbles were combined with
those clusters which have a radio core to form a second sub-sample.
Using $r_{\rm heat}=0.86r_{\rm cool}$ we calculate the size of an average bubble expected in these clusters.  In five
cases (3C129.1, A2063, A2204, A3112 and A3391) the
radio morphology is bi-lobed and its extent similar to the expected
bubble sizes.  A comparison
between the actual bubble size and the maximum expected if they were to offset the X-ray
cooling exactly, $R_{\rm max}$, shows a peak at $R_{\rm bubble}\sim 0.7 R_{\rm max}$ with a tail extending to larger $R_{\rm
bubble}/R_{\rm max}$.  The offset from the expected value of $R_{\rm
bubble}\sim R_{\rm max}$ may indicate the presence of a non-thermal
component in the innermost ICM of most clusters, with a pressure comparable to the thermal
pressure.
\end{abstract}

\begin{keywords}
galaxies: clusters: general -- galaxies: clusters: cooling flows --
X-rays: galaxies: clusters
\end{keywords}

\section{Introduction}

Since the discovery of ``holes'' in the X-ray emission at the centre
of the Perseus Cluster \citep{Bohringer}, and following the
launch of \emph{Chandra}, many more depressions
in the intra-cluster medium (ICM) of low redshift clusters have been
found (e.g. Hydra A, \citealp{McNamaraHydra00}; A2052,
\citealp{Blanton01}; A2199, \citealp{JohnstoneA2199}; Centaurus,
\citealp{SandersCent02}).  Recent compilations are given in \citet{DunnFabian05,
DunnFabian04, Birzan04}.  Such holes have been observed to
anti-correlate spectacularly with the radio emission from the active
galactic nucleus (AGN) at
the centres of these clusters.  Their morphology, particularly
in the closest clusters, has led to the interpretation that these are
bubbles of relativistic gas blown by the AGN into the thermal ICM.
This relativistic gas is less dense than the ICM and so the bubbles
are expected to detach from the core and rise up buoyantly through the
cluster, e.g. Perseus \citep{Churazov00,ACF_Halpha_PER03}.  The older,
detached, bubbles tend not to have
$\ghz$ radio emission associated with them and have been termed ``Ghost'' bubbles.

The X-ray emission of the ICM naturally leads to the conclusion that
the plasma should be cooling.  To maintain pressure
support, the gas is expected to flow on to the
central galaxy as a ``cooling flow.''  The subsequent increase in
density would lead to a ``cooling-catastrophe'', with extremely rapid
cooling in the cluster centre.  However, with the high spatial and
spectral resolution of \emph{Chandra} and \emph{XMM-Newton} little of
the expected X-ray emitting cool gas has
been found (\citealp{Peterson_2003}, see \citealp{PetersonFabian2006} for a review).  Many mechanisms have been proposed
by which the cool gas could be heated, including, for example, thermal conduction
\citep{KimNarayan03,VoigtFabian04}, but this appears not to work for clusters below
$5 \kev$.

A majority (71 per cent) of ``cooling core''  clusters harbour radio
sources \citep{Burns90}, and a similar fraction of clusters which
require heating (likely to be a cooling core) harbour clear bubbles
\citep{DunnFabian05}.  The action of creating the bubbles at the centre of the cluster by the
AGN is a favoured method of injecting energy into the central regions
of the cluster and so prevent the ICM from cooling.  This process sets
up sound/pressure waves in the ICM, the dissipation of which requires the ICM to be
viscous \citep{ACF_Halpha_PER03,ACF_Per_Mega_06}.  The viscous dissipation of the
pressure waves allows the energy from the bubble
creation to be dissipated far from the cluster centre, as gentle,
continuous and distributed heating as required \citep{VoigtFabian04}.
\citet{Birzan04} surveyed 80 clusters in the {\it Chandra} archive
without discriminating between cooling and non-cooling clusters and
found 16 that contained bubbles.  This was interpreted as clusters
having AGN/bubble activity only 20 per cent of the time.  In this work
we take a sample of clusters which forms a subset of an almost
complete sample and investigate the bubble population therein.  As the
parent sample is almost complete we are able to investigate prevalence
of cooling core clusters, duty cycles of bubble activity and radio sources within the general cluster
population.

The sample selection is described in Section \ref{selection},
and the data preparation and reduction in Section \ref{data_prep}.
The subset is itself then split into two: those clusters which are expected
to require some form of heating and harbour clear bubbles; and a
combination of those
which require some form of heating but do not harbour bubbles and those
clusters which have a central radio source.
 The analysis of those clusters with clear bubbles and of those without is
 discussed in Sections \ref{bubble_cluster} and \ref{nobubbles}
 respectively.  The overall implications of this work is discussed in
 Section \ref{discussion} with future investigations outlined in
 Section \ref{furtherwork}.

\section{Sample Selection}\label{selection}

\begin{table}
\centering
\caption{\label{Cluster_table} {\sc Cluster Sample}}
\begin{tabular}{llrc}
\hline
\hline
Cluster&Redshift&ObsId&Exposure (ks)\\
\hline
2A~0335+096&0.0349& 919&19.7\\
3C~129.1    &0.0223&2219&9.6\\
A85        &0.0521& 904&38.4\\
A262       &0.0164&2215&28.7\\
A478       &0.0882&1669&42.4\\
A496       &0.0223& 931&18.9\\
A1644      &0.0474&2206&18.7\\
A1651      &0.086 &4185&9.6\\
A1689      &0.181 &5004&19.9\\
A1795      &0.0627& 493&19.6\\
A2029      &0.0767& 891&19.8\\
A2052      &0.0348& 890&36.8\\
A2063      &0.0355&6263&16.8\\
A2199      &0.0300& 497&19.5\\
A2204      &0.1523& 499&10.1\\
A2597      &0.0824& 922&39.4\\
A3112      &0.0746&2516&16.9\\
A3391      &0.0545&4943&18.4\\
A3558      &0.0475&1646&14.4\\
A4059      &0.0478& 897&40.7\\
AWM7       &0.0172& 908&47.9\\
Centaurus  &0.0109&4954&89.1\\
Cygnus A   &0.0570& 360&34.7\\
Hydra A    &0.0522&4969&96.9\\
Klem44     &0.0283&4992&33.5\\
M87        &0.0037&2707&98.7\\
MKW3s      &0.0449& 900&57.3\\
Ophiuchus  &0.028 &3200&50.5\\
Perseus    &0.0183&3209&95.8\\
PKS~0745-191&0.1028& 508&28.0\\
\hline
\end{tabular}
\begin{quote} All the clusters in the sample, for the sample
  selection see text.  The exposure time is that after reprocessing
  the data.
\end{quote}
\end{table}

We use an updated version of the sample first outlined in \citet{DunnFabian05}, which is
a subset of the Brightest 55 (B55) sample.  The B55 sample was
compiled by \citet{Edge_1990} and studied in detail with {\it ROSAT}
data from a cooling-time point of view by \citet{Peres_1998}.  The B55
sample is a $2-10\kev$ flux-limited sample of
clusters of galaxies which are close enough to have been imaged by
previous X-ray satellites, and is nearly flux complete at high galactic
latitudes (all but nine are at $b>|20^|circ|$).

Out of the B55 sample, two clusters did not have  {\it
  ROSAT} observations, and four only had  {\it ROSAT HRI}
observations.  Out of the remainding 49 clusters,  \citet{DunnFabian05} selected those which had a
short central cooling time ($<3\gyr$) from the {\it ROSAT
  PSPC} and a large central temperature
drop ($T_{\rm outer}/T_{\rm centre}>2$), as these would be the clusters
requiring some form of heating so that large quantities of cool gas
are not formed.  Of the 49 clusters, 23 have a short $t_{\rm cool}$ and,
using the most recent temperature profiles, 21 have a large central
temperature drop.  20 clusters have both, of which 14 (70 per cent) show clear
depressions (bubbles) in the X-ray emission from the Intra-Cluster
Medium (ICM).  The duty cycle of bubbling in clusters which require
heating is therefore also at least 70 per cent.  From the whole sample
we find that the total fraction of clusters with clear bubbles from
current observations is at least 25 per cent (14/55).

Of the six clusters which require heating but have no clear bubbles, only one has no
radio source at the centre -- AWM7 \citep{Furusho04}.
2A 0335+096 has a complicated core structure, but there are depressions which have been interpreted as bubbles
\citep{Mazzotta_0335_03} and so this cluster is included with those which have
clear bubbles.  MKW3s also has a faint depression further out from the
centre, which has been interpreted as a bubble \citep{Mazzotta} and
this too is included with the clusters which have bubbles.

Also taken from the B55 sample are those clusters which are not
already in the sample and which host a radio core at the centre,
regardless of whether a {\it ROSAT} observation exists.  The
NVSS\footnote{NRAO (National Radio Astronomy Observatory) VLA (Very
  Large Array) Sky Survey} was used to find central radio sources.
The NVSS has extensive, but not complete sky coverage, and is not very
deep, so clusters which harbour faint central sources will be missed
\footnote{For example, \citet{Markovic_conf_04} present new detections of central sources in
A1650 and A2142 which are not detected in the NVSS.}. 
Out of the clusters not already in the sample, at least another 10 which have a central radio source in the NVSS.  So the fraction of clusters
which have a central radio source is at least $\sim53$ per cent (29/55, i.e. including those with bubbles).  These two cluster samples contain
a total of 30
clusters (see Table \ref{Cluster_table}), 16 of which have a radio
source, clear bubbles and require heating\footnote{\citet{Donahue_05} investigate both A1650
  and A2244 as they have radio-quiet cooling cores ($t_{\rm cool}<H_0^{-1}$).  
  A2244 has $t_{\rm cool}>3\gyr$ from \citet{Peres_1998} and athough \citet{Markovic_conf_04} detect a radio source
  at the centre of A1650 none is seen in the NVSS and so neither is in
  our sample.}, 13
which have a radio source but no bubbles, and AWM7 which
has no evidence for radio activity.  The assignment of the clusters to
the different groups is shown in Table \ref{Cluster_assign}.

We have chosen a central cooling time of $t_{\rm cool}<3\gyr$ to investigate those clusters
where a cooling flow would form if there is no heating.
These are therefore the clusters in which extreme heating rates are expected.  Changing to $t_{\rm
  cool}<7\gyr\sim t_{\rm Hubble}/2$ could add another nine clusters into the sample which
require heating. There is only one cluster of these nine which also has a
central temperature drop -- A2142 -- and so the fraction of clusters
with bubbles from those which require heating changes little  Three of these nine clusters (A2063, A1689 \& A1651) already
fall into the sample which have core radio detections.  None has a
report of clear bubbles.

\begin{table}
\centering
\caption{\label{Cluster_assign} {\sc Cluster Sub-samples}}
\begin{tabular}{l|ll|l}
\hline
\hline
Clear Bubbles&\multicolumn{2}{c}{NVSS Radio}&No Radio\\
\hline
Heating&Heating&No Heating&Heating\\
\hline
A85	&2A~0335+096$^\dagger$	&3C~129.1	&AWM7\\
A262	&A496			&A1644	&\\
A478	&A2204			&A1651	&\\
A1795	&PKS~0745-191		&A1689	&\\
A2029	&MKW3s$^\dagger$	&A2063	&\\
A2052	&			&A3112	&\\
A2199	&			&A3391	&\\
A2597	&			&A3558	&\\
A4059	&			&Klem44	&\\
Centaurus&			&Ophiuchus	&\\
Cygnus A&			&	&\\
Hydra A	&			&	&\\
M87	&			&	&\\
Perseus	&			&	&\\
\hline
14&5&10&1\\
\hline
\end{tabular}
\begin{quote}{\scshape Notes:}$^\dagger$ These clusters have
  complicated cores/faint depressions and are added to the Clear
  Bubbles set (for further discussion see text).  The Clear Bubbles
  (incl $^\dagger$) clusters form one subset and the remainder form the other.
\end{quote}
\end{table}

\subsection{Heating Distribution}

\begin{figure}
\centering
\includegraphics[width=1.0 \columnwidth]{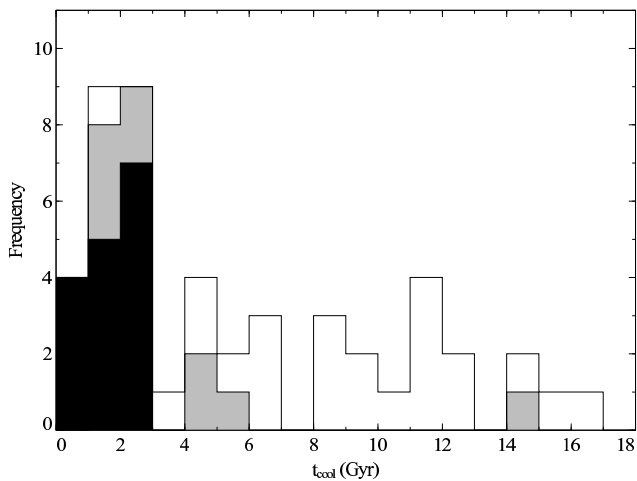}
\caption{\label{t_cool_hist} The distribution of $t_{\rm cool}$ in the
  B55 sample.  The black bars indicate those clusters with clear
  bubbles, the grey are for the clusters with no bubbles but with
  radio sources.  This graph has been truncated, two clusters with
  $t_{\rm cool}=22$ and $32\gyr$ (A1367 \& A2255) have been removed to
expand the $x$-axis.}
\end{figure}

The distribution of the central cooling times of the B55 clusters is shown
in Fig. \ref{t_cool_hist}.  Almost all of the clusters which have a
cooling time less than $3\gyr$ contain a radio source or clear
bubbles, with the exception being AWM7.  The clusters in the B55 sample
which do not have central cooling time values from the {\it ROSAT
  PSPC} have not been included in the figure.

\section{Data Preparation}\label{data_prep}

The X-ray {\it Chandra} data of the clusters were
processed and cleaned using the CIAO software and calibration files
(CIAO v3.3, CALDB v3.2).
We began the reprocessing by removing the afterglow detection and
re-identifying the hot pixels and cosmic ray afterglows, followed by
the tool {\scshape acis\_process\_events} to remove the pixel randomisation and to
flag potential background events for data observed in Very Faint (VF)
mode.  The Charge-Transfer Efficiency was corrected for, followed by
standard grade selection.  Point-sources were identified using the
{\scshape wavdetect} wavelet-transform procedure.  For clusters observed with
the ACIS-S3 chip, the ACIS-S1 chip was used to form the light curves
where possible.  In all other cases, light-curves were
taken from on-chip regions as free as possible from cluster emission.
For the spectral analysis, backgrounds were taken from the CALDB
blank-field data-sets.  They
had the same reprocessing applied, and were reprojected to the correct
orientation.  

Cluster centroids were chosen to lie on the peak in the X-ray
surface brightness.  
Annular regions were automatically assigned with constant
signal-to-noise, stopping where the calculated surface brightness of the cluster dropped
below zero.   The initial signal-to-noise was 100, and this was
increased or decreased by successive factors of $\sqrt{2}$ to obtain a number of regions between four and
ten.  The minimum signal-to-noise allowed was 10.

The $0.5-7\kev$
spectra were extracted, binned with a minimum of 20 counts per bin,
and, using {\scshape xspec} (v12.2.1r)
(e.g. \citealt{Arnaud_xspec_96}), a {\scshape projct} single temperature
{\scshape mekal} (e.g. \citealt{Mewe_mekal_95}) model with a {\scshape phabs}
absorption was used to
deproject the cluster.  In some clusters the temperatures
for some of the regions were undefined.  To solve this problem, the minimum number of regions
was reduced, a maximum radius for the outermost annulus was set, or the outermost region was
removed to try to improve the behaviour of the profile (this was
required for  3C129.1, A496, A1689, A2063, A2597, A3391, A4059,
Klem 44, Ophiuchus \& PKS0745).  Using the deprojected cluster temperature, abundance and normalisation profiles; density, pressure,
entropy, cooling time and heating profiles were created.  The
temperature profiles and the derived profiles for the clusters with
clear bubbles are shown in the Appendix
(Figs. \ref{combined_temp} to \ref{combined_heating_kpc}).

These profiles give aziumthally averaged values for the cluster
properties and have been used in the subsequent calculations.  In some
clusters, notably M87 and Perseus, the central parts of the cluster
are not very smooth, e.g. due to bubbles.  \citet{Donahue06} show that
these features do not strongly bias estimates of the entropy.
As such the use of these azimuthally averaged values is not likely to
introduce large biases into the subsequent calculations.

\section{Clusters with Bubbles}\label{bubble_cluster}

To estimate the energy input of the AGN to the ICM we estimated the
energy required to create the observed bubbles.  If the expansion
rates are slow then this is the sum of the bubble's internal energy
and the $P_{\rm th}{\rm d}V$ work done.
\begin{equation}
E=\frac{1}{\gamma_1-1}P_{\rm th}V + P_{\rm th}V=\frac{\gamma_1}{\gamma_1-1}P_{\rm th}V
\end{equation}
where $V$ is the volume of the bubble, $P_{\rm th}$ is the thermal pressure of the surrounding ICM and
$\gamma_1$ is the mean adiabatic index of the fluid in the bubble.  In
the case where the fluid in the bubble is relativistic, $\gamma_1=4/3$,
and so the total energy in the bubble is $4P_{\rm th}V$.  As the bubbles are
approximately elliptical we parameterise with a semi-major axis
along the jet direction, $R_{\rm l}$, and a semi-major axis across it
$R_{\rm w}$.  The volumes are therefore $4\pi R_{\rm l}R_{\rm w}^2/3$,
where we have assumed that the bubbles are prolate ellipsoids.  

To obtain an estimate for the rate at which energy from a bubble would be
dissipated in the ICM, the age for the bubble, $t_{\rm age}$, is required.  Most of
the bubbles in the sample are young and still attached to their radio core.  In
these cases the timescale used is the sound speed timescale ($t_{\rm age}=t_{\rm
  c_s}=R/c_{\rm s}$) where the
sound speed is given by
\begin{equation}
c_{\rm s} = \sqrt{\gamma_2 k_{\rm B}  T/\mu m_{\rm H}} 
\end{equation}
where $\gamma_2=5/3$ for a non-relativistic gas.
This timescale arises from the fact that no strong shocks have been observed in
clusters.  From this the bubbles are assumed to be expanding at not
much more than the local sound speed of the ICM.  In the cases where
the bubbles are ghost bubbles, i.e. they do not contain $\ghz$ radio
emission and are detached from the cluster centre, we have used the
buoyancy timescale ($t_{\rm age}=t_{\rm
  b}=R/v_{\rm b}$) as these are expected to be rising buoyantly up
through the ICM, with the buoyancy velocity given by
\begin{equation}
v_{\rm b}=\sqrt{2gV/SC_{\rm D}}
\end{equation} 
where $C_{\rm D}=0.75$ is the drag coefficient \citep{Churazov01} and $S=\pi r_{\rm w}^2$ is the
cross-sectional area of the bubble.  Those bubbles which have been classed as
Ghost bubbles are indicated in Table \ref{Bubble_results}
For further discussion on the timescales see Section
\ref{timescales} and \citet{DunnFabian05,DunnFabian04}.  The energies and timescales for the
bubbles were calculated from values for the radii in
\citet{Allen_Bondi_06,DunnFabian05, DunnFabian04}, except
for 2A 0335+096 which are from \citet{Birzan04}.

From the density profiles and the relevant cooling function, the heat
input required per spherical shell was calculated.  By comparing this to the
total power provided by the bubble ($P_{\rm bubble}=4P_{\rm th}V/t_{\rm age}$), the radius
out to which the bubble power can offset the X-ray cooling was
calculated for each cluster ($r_{\rm heat}$, see Table \ref{Bubble_results}).  The
radius is quoted in $\kpc$ and also as a fraction of $r_{\rm cool}$ (for a cooling time of $3\gyr$).  For an
estimate on the uncertainties, Monte-Carlo simulations of the
calculations were performed.

\begin{table*}
\centering
\caption{\label{Bubble_results} {\sc Bubble Heating}}
\begin{tabular}{lllcccccccc}
\hline
\hline
Cluster&Bubble&Ghost?&$r_{\rm l}$&$r_{\rm w}$&$R_{\rm dist}$&$L_{\rm cool}^{\hspace{0.35cm} a}$&$r_{\rm cool}$&$r_{\rm heat}$&$r_{\rm heat}/r_{\rm cool}^{\hspace{0.35cm} b}$&Heated Fraction$^{c}$\\
&&&$(\kpc)$&$(\kpc)$&$(\kpc)$&$10^{43}\ergps$&$(\kpc)$&$(\kpc)$&&\\
\hline

    2A~0335 &E&G&9.3&6.3&23& $    22.7\pm    0.4$ & $	69.8\pm    1.0$ & $  14.8\pm	1.62$ & $   0.21\pm   0.02$ & $   0.09\pm   0.02$ \\ 
&W&G&4.8&2.6&28&\\
       A85 &N&G&5.3&7.0&14& $      9.0\pm    0.4$ & $	47.6\pm    1.3$ & $  81.4\pm  10.7  $ & $   1.71\pm   0.23$ & $   2.54\pm   0.71$ \\ 
&S&G&6.4&8.6&22&\\
      A262 &E&&2.4&1.8&3.2& $      0.5\pm    0.0$ & $	31.4\pm    0.8$ & $  28.9\pm	3.20$ & $   0.92\pm   0.10$ & $   0.93\pm   0.10$ \\ 
&W&&2.7&1.9&3.4&\\
      A478 &NE&G&4.0&6.9&9.8& $   78.2\pm    3.0$ & $	84.8\pm    2.7$ & $  18.8\pm	2.63$ & $   0.22\pm   0.03$ & $   0.13\pm   0.02$ \\ 
&SW&G&4.5&7.7&10.7&\\
     A1795 &NW&&4.1&3.1&3.8& $    26.9\pm    2.1$ & $	68.8\pm    3.3$ & $  38.1\pm	1.63$ & $   0.55\pm   0.04$ & $   0.35\pm   0.04$ \\ 
&S&&4.8&2.8&5.2&\\
     A2029 &NW&&7.2&2.2&9.4& $    51.2\pm    9.4$ & $	73.3\pm    9.9$ & $  31.9\pm	3.50$ & $   0.43\pm   0.08$ & $   0.31\pm   0.08$ \\ 
&SE&&6.5&2.9&9.4&\\
{\bf A2052} &N&&10&10&10& $        4.5\pm    0.1$ & $	46.2\pm    1.0$ & $ 226  \pm	0.00$ & $   4.90\pm   0.11$ & $   6.83\pm   0.83$ \\ 
&S&&15&15&15&\\
     A2199 &E&&2.3&1.3&2.3& $      5.3\pm    0.3$ & $	46.0\pm    1.1$ & $  16.7\pm	0.92$ & $   0.36\pm   0.02$ & $   0.20\pm   0.03$ \\ 
&W&&2.4&1.4&2.4&\\
     A2597 &NE&G&7.8&7.8&21& $    33.0\pm    1.5$ & $	66.6\pm    1.0$ & $  63.2\pm	5.08$ & $   0.95\pm   0.08$ & $   0.85\pm   0.14$ \\ 
&SW&G&12&7.8&25&\\
     A4059 &N&&5.4&2.2&5.4& $    2.4\pm    0.2$&  $   45.5\pm    2.0$ & $  46.91\pm   3.22$ & $   1.03\pm   0.08$ & $   1.06\pm   0.18$ \\ 
&S&&3.9&4.8&3.9&\\
 Centaurus &E&&2.8&1.7&2.5& $      2.1\pm    0.0$ & $	42.8\pm    0.2$ & $  20.7\pm	2.99$ & $   0.48\pm   0.07$ & $   0.44\pm   0.07$ \\ 
&W&&2.7&2.0&3.0&\\
{\bf Cygnus A} &E&&30&21&44& $    29.8\pm    1.1$ & $	55.1\pm    1.6$ & $ 359  \pm	0.00$ & $   6.52\pm   0.19$ & $  38.0 \pm   5.8 $ \\ 
&W&&34&22&48&\\
Hydra$^{\rm d}$ &N&&20&5.5&18& $  15.6\pm    0.2$ & $	56.2\pm    0.7$ & $  83.5\pm  11.6  $ & $   1.49\pm   0.21$ & $   1.42\pm   0.14$ \\ 
&S&&17&4.4&22&\\
{\bf  M87}  &E&&1.5&1.5&1.5& $     2.5\pm    0.0$ & $	27.4\pm    0.0$ & $  27.4\pm	0.00$ & $   1.00\pm   0.00$ & $   0.66\pm   0.02$ \\ 
&W&&1.9&1.2&1.9&\\
     MKW3s &S&G&13&21&56& $        5.0\pm    0.3$ & $	53.4\pm    2.0$ & $ 110  \pm  29.7  $ & $   2.06\pm   0.56$ & $   2.48\pm   0.67$ \\ 

{\bf Perseus} &N&&8.2&8.2&8.2& $  50.2\pm    0.1$ & $	76.4\pm    0.1$ & $ 109  \pm	0.00$ & $   1.43\pm   0.00$ & $   2.29\pm   0.03$ \\ 
&S&&8.9&8.9&8.9&\\		  

\hline
     M87$^e$ &E&&1.5&1.5&1.5&  $    1.5\pm    0.1$ & $   37.4\pm    2.3$ & $  35.3\pm   3.0$ & $   0.94\pm   0.10$ & $   0.93\pm   0.15$ \\
&W&&1.9&1.2&1.9& \\
\hline
\end{tabular}
\begin{quote}
The clusters in bold are those where the bubbles heat beyond the
analysed region.  For M87 the cooling
radius could not be determined from a {\it ACIS} single chip.  These clusters
were not included in the calculation of the mean values.

$^{a}$ $L_{\rm cool}$ calculated for the $0.5-7.0\kev$ range.
$^{b}$ The radius, as a fraction of the cooling radius, out to
which the bubble power ($4P_{\rm th}V/t_{\rm age}$) can offset the X-ray
radio source by interacting with its surroundings cooling.
$^{c}$ The fraction of the X-ray cooling that occurs within the
cooling radius which is offset by the energy of the bubble's expansion.
$^{d}$ The bubble sizes for Hydra A are not those from the cluster
scale outburst from \citet{NulsenHydra04} but from the smaller, inner radio emission.
$^{e}$ These values for M87 come from the data in \citet{Ghizzardi04}
which extend to larger radii.

\end{quote}
\end{table*}

The bubbles in A2052, Cygnus A and
Perseus supply enough energy to offset all the X-ray cooling far beyond the cooling radius.
In fact the energy they supply offsets all the X-ray cooling in the
analysed regions of the cluster.  In the Perseus cluster this is mainly because the cluster is so close, and so only
the innermost regions fit onto one chip.  In A2052 and Cygnus A the
surface brightness of the X-ray emission falls to zero before the edge
of the chip.  As a result the spectral analysis stops at a radius such
that the bubble energy offsets more than all the X-ray cooling within
that radius.  In the case of M87, the cluster is so close that
the cooling radius, $r_{\rm cool}$, could not be calculated from a single chip; however the bubble energy
is greater than the X-ray cooling within the largest radius obtained on a
single chip.  These clusters are highlighted in Table
\ref{Bubble_results}.   The results for M87 obtained from
 \citet{Ghizzardi04} are shown for comparison and indicate that the
current jet and counter jet cavities approximately offset all the
X-ray cooling within $r_{\rm cool}$.  This result, however, has not been
included in any of the subsequent analysis except where explicitly stated.

\begin{figure}
\centering
\includegraphics[width=1.0\columnwidth]{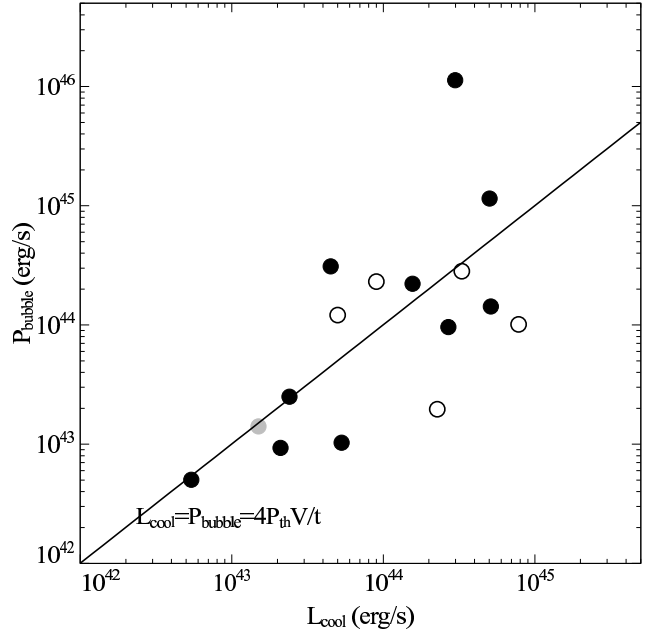}
\caption{\label{L_vs_P} The X-ray Luminosity within the cooling radius
versus the bubble power for the clusters.  Clusters that fall above
the line have AGN which inject sufficient energy into the ICM to
offset all of the X-ray cooling.  The four clusters with open circles
are those with Ghost bubbles (2A 0335, A2597, A85, A478 and MKW
3s). The grey point is M87 using the data from \citet{Ghizzardi04}.}
\end{figure}

\begin{figure}
\centering
\includegraphics[width=1.0\columnwidth]{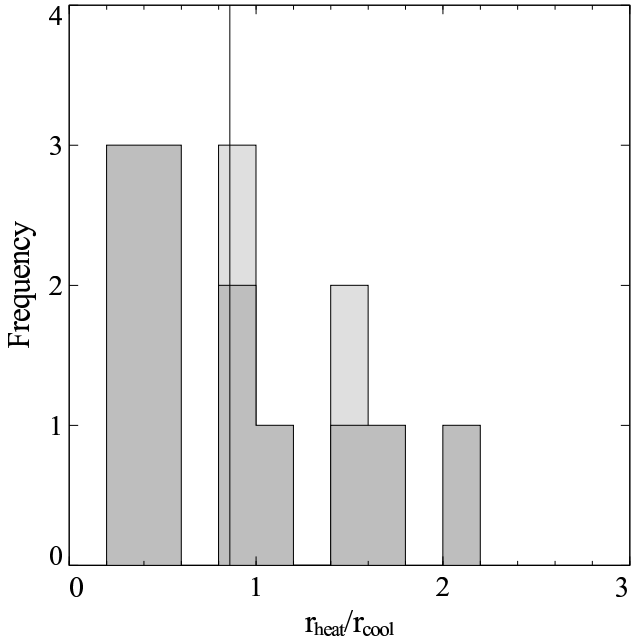}
\caption{\label{binnedheating} The distribution of the radius (as a
  fraction of the cooling radius) out to which the energy in the
  bubbles can offset the X-ray cooling.  The extra points at
  $r_{\rm heat}/r_{\rm cool}=0.8-1$ and $r_{\rm heat}/r_{\rm cool}=1.4-1.6$ are M87 and
  Perseus respectively.  A2052 and Cygnus A are not
  included on this plot. All four of these clusters were not taken
  into account when calculating the mean, indicated by the vertical line.}
\end{figure}

The average distance, as a fraction of the cooling radius, out to
which the bubbles can offset the X-ray cooling is estimated using the
mean of the results in Table \ref{Bubble_results}.  The
uncertainties in the mean were estimated using a simple
bootstrapping method.  As a fraction of the cooling radius the mean distance out to
which the bubbles offset the X-ray cooling is $r_{\rm heat}/r_{\rm
  cool}=0.86\pm0.11$.   The mean
power that the bubbles supply, as a fraction of the X-ray cooling within
the cooling radius, is $0.89\pm0.16$.  The distribution of the distance out to
which the bubbles offset the X-ray cooling is shown in
Fig. \ref{binnedheating}.  The clusters where the bubble power is such
that it is greater than the all X-ray cooling within the analysed region
(A2052, Cygnus A \& Perseus) and M87 have not been included in the estimate on the
mean.  Adding A2052, Cygnus A and Perseus into the estimation of the mean gives
$r_{\rm heat}/r_{\rm cool}=1.45\pm0.29$ (with a heated fraction of $3.49\pm1.93$), and so it is likely that the value of
$r_{\rm heat}/r_{\rm cool}=0.86$ is a lower limit for this sample of
clusters.  See Section \ref{discussion} for further discussion of the
range of $r_{\rm heat}/r_{\rm cool}$ and its implications.  Comparing $L_{\rm cool}$ with $P_{\rm bubble}$
(Fig. \ref{L_vs_P}) shows that there is a slight correlation between
the cooling luminosity and the bubble power.  The clusters lying above
have AGN which supply sufficient energy into the ICM to offset the
cooling.  

In a set of clusters harbouring clear bubbles \citet{Birzan04} and
\citet{Rafferty06} found that the mechanical luminosities required to
offset the cooling ranged between $1P_{\rm th}V$ and $20P_{\rm th}V$. Around half the
objects in their sample had cavities which (assuming $4P_{\rm th}V$) could
offset the cooling though there is a large spread, with some objects
falling short.  In this sample, which is drawn from an almost
flux-complete parent sample, we find similar results.

As yet the method by which the bubbles dissipate their energy into the
ICM has not been clearly determined.  The energy also needs to be
transported out to a large radius, for example, the viscous
dissipation of sound waves in the ICM as they travel out in the
cluster \citep{ACF_deep_PER03,Fabian_05_visc_cond}.  As these are seen at large
radii, they can plausibly carry their energy sufficiently far out.  As
a result, the bubbles analysed here, on average, supply enough energy
to offset the X-ray cooling, but whether this energy actually performs
this function is currently unknown.

\section{Clusters without Bubbles}\label{nobubbles}

The second cluster subset is a combination of those clusters which
require heating and show {\it no}
clear bubbles, and those clusters which harbour a radio source (see
Table \ref{Cluster_assign}).  The amount of
X-ray cooling within $r_{\rm heat}/r_{\rm cool}=0.86$ (the mean distance out to
which the bubbles offset the X-ray cooling for the other cluster subset) for these clusters was
calculated.  In the assumption that any X-ray cooling from these
clusters is also offset by AGN bubbles, and that these bubbles are young and so expanding at the sound speed, the
dimensions of these bubbles was calculated.  The resulting dimensions
are shown in Table \ref{Hidden_bubbles}, where we have assumed that
bubbles occur in pairs.  The uncertainties have been
estimated using simple Monte-Carlo simulations of the calculation.

The total X-ray counts expected within the area of the predicted bubble in the
current observation was
calculated from the average X-ray counts per pixel in the
central regions of the cluster.  The central regions were chosen so
that it could be reasonable assumed that the surface brightness is
constant.  This is, however, a lower limit on the number of counts
and the signal to noise for the innermost regions, i.e. those where any
bubbles are likely to be.  This gave the expected X-ray signal
to noise which is shown in Table \ref{Hidden_bubbles}.  Using the
bubbles in A2052 and Hydra A as templates, we estimate that the counts
in the centre of the bubbles are around 30 percent lower than the
counts in the rims.  

So that any bubbles could be detected above the noise
at a $3\sigma$ level the X-ray signal to noise has to be greater than around
10 (equivalent to noise at 10 per cent level).  This does, however,
rely on knowing where the bubbles are in the centre of the cluster.
If there is extended radio emission then the average counts interior
and exterior to the expected bubble can be compared to see if there is
a net decrement within the bubble.  Half of the clusters in the sample
have signal-to-noise values of less than 10, and so, even if extended radio
 emission is detected, no significant detection of
bubbles from the current X-ray observations are likely.  The counts
per pixel in the central regions of these clusters is also very low.
 
A496, A2204, A3112, AWM7, Klem44, Ophiuchus and PKS~0745 all have X-ray
signal-to-noise which is such that if a bubble of the expected size
were present it should be detectable in the current X-ray
observation.  This is, however, still with the caveat that the size
and location of the bubble is known (again, extended radio emission
would help with this).

To be able to detect a decrement in the X-ray emission without any
radio emission
to guide the eye is more difficult.  AWM7 and Klem44 only have around
one count per pixel in the central regions, and so it is unlikely that
a bubble could be identified by eye from the X-rays alone.  It is more likely that in
clusters with higher counts per pixel (A2204, Ophiuchus and PKS~0745)
that bubbles would be detectable as these have more than 1000 counts
expected within the area of the expected bubble.  

The contrast ratio we have chosen, however, is for very
clear bubbles with well defined rims in nearby clusters.  In smaller
radio sources, bubbles may be less
clearly defined, and so the X-ray signal-to-noise may have to be much
higher to be able to clearly detect cavities if they are there.  

\begin{table*}
\centering
\caption{\label{Hidden_bubbles}{\sc Predicted Bubble Sizes}}
\begin{tabular}{lcccrr}
\hline
\hline
Cluster&$r_{\rm cool}$&\multicolumn{2}{c}{Predicted Bubble
  Radius}&X-ray S/N&counts/pix\\
&$(\kpc)$&$(\kpc)$&arcsec&\\
\hline
  3~C129.1 & $   5.89\pm   3.08$ & $   0.60\pm   0.16$ & $   1.34\pm   0.36$ &     0.9  &    0.04 \\ 
      A496 & $  54.83\pm   1.97$ & $   5.50\pm   0.53$ & $   8.40\pm   0.80$ &    53.7  &    3.15 \\ 
     A1644 & $  24.55\pm   9.05$ & $   1.98\pm   0.34$ & $   2.13\pm   0.36$ &     4.6  &    0.37 \\ 
     A1651 & $  14.24\pm  12.53$ & $   1.40\pm   0.49$ & $   0.87\pm   0.30$ &     1.5  &    0.22 \\ 
     A1689 & $  63.12\pm   5.85$ & $   5.19\pm   0.42$ & $   1.70\pm   0.14$ &     7.0  &    1.32 \\ 
     A2063 & $  29.27\pm   1.57$ & $   2.09\pm   0.38$ & $   2.95\pm   0.53$ &     6.5  &    0.37 \\ 
     A2204 & $  69.44\pm   3.16$ & $   8.13\pm   1.04$ & $   3.06\pm   0.39$ &    34.1  &    9.56 \\ 
     A3112 & $  57.86\pm   2.71$ & $   5.44\pm   0.62$ & $   4.05\pm   0.46$ &    30.4  &    4.34 \\ 
     A3391 & $  10.84\pm   2.18$ & $   2.61\pm   0.33$ & $   2.52\pm   0.32$ &     2.5  &    0.08 \\ 
     A3558 & $  14.90\pm   1.46$ & $   1.60\pm   0.10$ & $   1.72\pm   0.10$ &     4.0  &    0.41 \\ 
      AWM7 & $  24.84\pm   2.34$ & $   1.25\pm   0.22$ & $   3.58\pm   0.63$ &    15.6  &    1.46 \\ 
    Klem44 & $  38.79\pm   1.41$ & $   2.56\pm   0.21$ & $   4.61\pm   0.39$ &    16.1  &    0.94 \\ 
      Ophi & $  32.85\pm   0.82$ & $   1.69\pm   0.32$ & $   3.01\pm   0.57$ &    28.6  &    6.93 \\ 
  PKS~0745 & $  90.76\pm   2.16$ & $  10.08\pm   2.03$ & $   5.32\pm   1.07$ &    51.9  &    7.31 \\ 
\hline
\end{tabular}
\end{table*}

Using the Very Large Array (VLA)\footnote{The
National Radio Astronomy Observatory is operated by Associated
Universities, Inc., under cooperative agreement with the National
Science Foundation.} and Australia Telescope Compact
Array (ATCA)\footnote{The Compact Array is part of the Australia
  Telescope, which is funded by the Commonwealth of Australia for
  operation as a National Facility managed by CSIRO.} archives we attempted to find observations of
these clusters to investigate the morphology of the radio sources at
their centres.  In most cases the radio sources were not resolved in
the archival observations.  However for at least four of the clusters in Table \ref{Hidden_bubbles} there
are extended radio sources which could be bubbles that
have not been detected in the X-ray images of clusters.  In other
cases it has not been possible to find radio data of high enough
resolution, and so there may be further clusters with, as yet, undetected
bubbles.

\begin{figure}
\centering
\includegraphics[width=1.0\columnwidth]{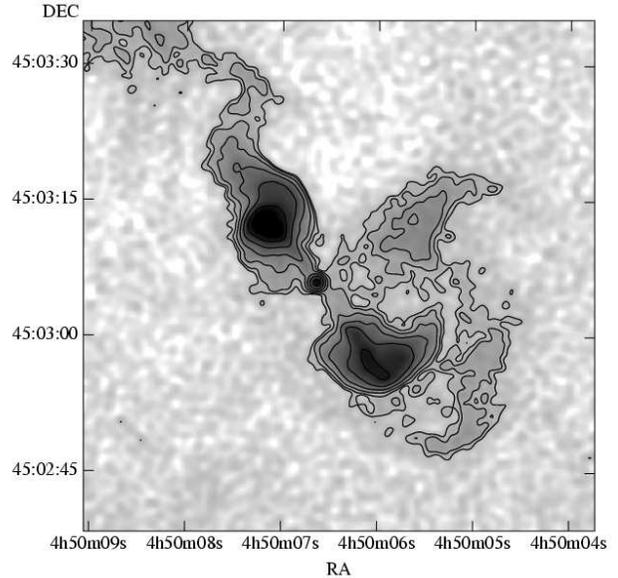}
\caption{\label{3c129} The $5 \ghz$ radio emission from 3C129.1 \citep{Taylor3C129}.}
\end{figure}

In the case of 3C~129.1 the morphology of the radio emission is very
suggestive that the surrounding ICM is constraining the expansion of
the inner radio lobes (see Fig. \ref{3c129}).  The X-ray
observations to date, however, show no clear interactions of the radio source
with the ICM.  The observed radio lobes in 3C~129.1 are
$\sim 2.6\kpc$ $(5.5\arcsec)$ in radius, which is around four times that expected for an
``average'' bubble in this cluster, $0.60\pm0.16\kpc$ $(1.34\pm0.36
\arcsec)$.  The observed radio lobes correspond to around 10 {\it Chandra}
pixels in radius (which are $0.49 \times 0.49\arcsec$), and so if
the radio lobes are excavating cavities in the ICM, then the resultant
bubbles should be seen in deep X-ray observations of the cluster which
would increase the X-ray signal-to-noise.  

A3391 also harbours an extended radio source with a bi-lobed
morphology \citep{OtaniA3391,MorgantiA3391}.  The radio lobes in this
cluster are separate from the radio core.  If these lobes have formed
cavities in the ICM, this separation implies that the cavities
are older bubbles which have detached from the cluster centre and are
rising buoyantly.  The radio
structure is not smooth and simple, but the $13\cm$
($2.38\ghz$) observations
from \citet{OtaniA3391} show the lobes as $\sim\!8\times12\kpc$
($8\times12\arcsec$ or $16\times24$ {\it Chandra} pixels) in size.  The
expected size of an average bubble is $2.61\pm0.33\kpc$, which is much
smaller than the observed size of the radio emission.  

The {\it ROSAT} X-ray
maps of the centre of this cluster show no clear indication that the
radio source is interacting with its surroundings.  An explanation for
  these large detached radio lobes is if close feedback exists between the X-ray
cooling and the AGN heating, then there may recently have been a past cycle
  of AGN heating and now there is a period of quiescence and no (small) lobes are seen with the expected sizes.  

The central radio source in A2063 also has two peaks of emission.  The
dimensions of the observed radio emission are $\sim\!
2\times1\arcsec$ corresponding to $\sim\!1.4\times0.7 \kpc$ (or $4\times2$ {\it Chandra}
pixels).
The expected sizes for average radio bubbles in this cluster are
$2.09\pm0.38\kpc$ ($2.95\pm0.53\arcsec$ or $\sim 6$ {\it Chandra}
pixels) in size.

None of the X-ray observations of these three clusters are deep enough
so that clear indications of an interaction between the radio source
and the ICM are expected.

The 1.32~GHz radio map of A3112 shows two faint diffuse regions of emission to
the south east and south west of the core \citep{Takizawa2003}.
Unfortunately the radio image has an elogated beam which makes it
difficult to determine whether a faint excess in the X-rays is
definitely associated with the radio lobes.  The observed radio
emission is $\sim\!4\arcsec$ in radius corresponding to $\sim\!5\kpc$ (or $ \sim8 $ {\it Chandra} pixels).  The emission is offset
from the core, so it may correspond to detached bubbles which still
have some GHz radio emission.  The expected sizes for average radio
bubbles in this cluster match almost exactly the observed radio emission.

\begin{figure}
\centering
\includegraphics[width=0.98\columnwidth]{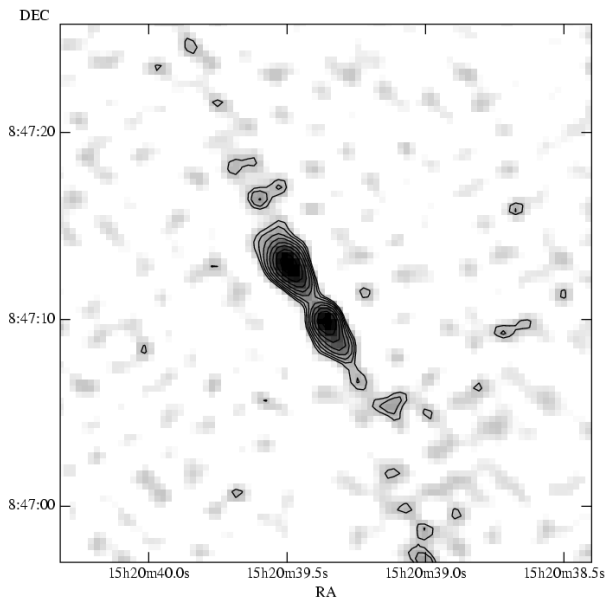}
\caption{\label{A2063} The $4.9 \ghz$ radio emission from A2063.}
\end{figure}

\begin{figure}
\centering
\includegraphics[width=1.0\columnwidth]{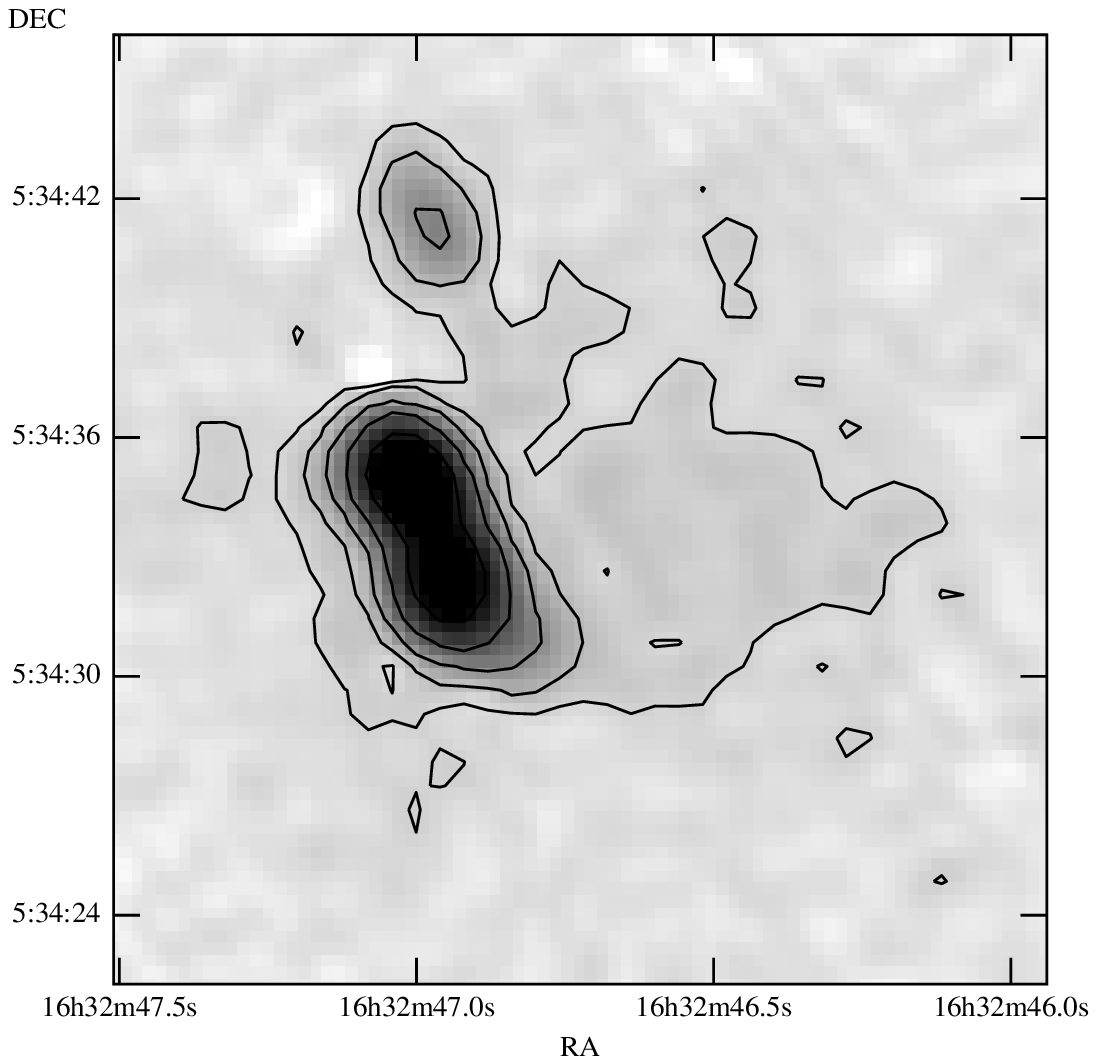}
\caption{\label{A2204} The $1.5 \ghz$ radio emission from A2204 \citep{Sanders_A2204}.}
\end{figure}

A2204 and PKS 0745 are the two of the more distant clusters in our sample,
with redshifts greater than $0.1$ ($0.1523$ and $0.1028$ respectively).  The
radio morphology of A2204 shows two clear maxima (see Fig. \ref{A2204}), which could be the
two lobes of the radio source, which have not been fully resolved.
Again, the X-ray emission shows no clear indication of any
interaction of the radio source with the ICM.  The radii of the
observed radio maxima in A2204 are $\sim5\kpc$ ($\sim2\arcsec$ or 4
{\it Chandra} pixels), and the size
of the average bubble expected in this cluster is only slightly larger
($8.13\pm1.04\kpc$ or $3.06\pm0.39\arcsec$).  

PKS 0745 has an amorphous radio source at the cluster centre, rather than a
clear bi-lobed morphology and it is unclear as to the effect of the
central radio source on the ICM.  The {\it total} dimensions of the observed radio
source are $9.2\times6.0\kpc$, which is not that dissimilar from that
for an expected bubble ($r=10.1\pm2.0\kpc$).

The X-ray observations pf these three clusters are sufficiently deep that if
any bubbles existed and the surface brightness contrast between the
rims and the bubble centre were only 10 per cent they should be
observed, especially as there is extended radio emission to guide the
eye.  It is possible that the contrast between the bubbles and the
surrounding rims in these two clusters is very low, and as a result
any cavities have not yet been detected.  In any case, deeper X-ray
observations would allow further investigation into the presence or
absence of bubbles in these clusters.

The current X-ray observations for most of these clusters are too
short to provide sufficient signal-to-noise for any cavities to be
clearly identified in the images.  As a result some cluster may
harbour bubbles which cannot be detected.  Some of the clusters do,
however, harbour extended radio sources whose dimensions are not
dissimilar to those expected from an ``average'' bubble, should one be
present in the cluster.

\section{Discussion}\label{discussion}

We first discuss the spread in values of $r/r_{\rm cool}$ and then
investigate the possibility of the young bubbles being below their
maximum size.  Finally we detail some of the uncertainties and biases
present in this analysis.

Although the mean of the radius out to which the
bubbles can offset the X-ray cooling is $0.86r_{\rm cool}$ (from bootstrapping), there is a large spread
in the values (see Fig. \ref{binnedheating}).  In some clusters the bubble power offsets the heating
out to over $1.5r_{\rm cool}$ (Hydra) whereas in others it only reaches
out to $0.36r_{\rm cool}$ (A2199).  The range of the energy supplied
by the bubbles as a fraction of that required within $r_{\rm cool}$ also has a large
spread, from $\times1.4$ that required in Hydra, to $\times0.20$ in
A2199, with an average of $0.89$.  The Ghost bubbles
are not taken into account here, along with the clusters where the analysis
is not complete (A2052, Cygnus A and Perseus, but not M87).  Adding these in the
range increases from $0.22$ to $6.5r_{\rm cool}$.  So in some
clusters the bubbles can provide (more than) enough heating to offset
the X-ray cooling, but some fall short.  

\begin{figure}\centering

\includegraphics[width=1.0\columnwidth]{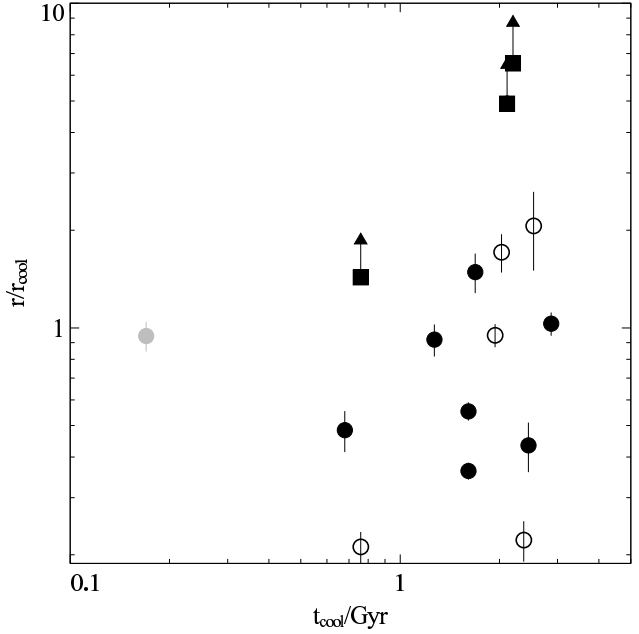}

\caption{\label{rcool_tcool} The distance out to which the bubbles can
  offset the X-ray cooling versus the central cooling time.  The three
clusters which are have lower limits are Perseus, A2052 and Cygnus A
(in increasing $t_{\rm cool}$).  The four clusters with open circles
are those with Ghost bubbles (2A 0335, A2597, A85, A478 and MKW
3s). The grey point is M87.}
\end{figure}

Fig. \ref{rcool_tcool} shows the distance to which the bubbles can
offset the X-ray cooling versus the central cooling time (from
\citealt{Peres_1998} adjusted to our cosmology).  The grey
point is M87 using the data presented in \citet{Ghizzardi04}.  The
three squares are A2052,
Cygnus A and Perseus where the bubbles provide energy to beyond where it was
possible to deproject the cluster.  Unfortunately there is no clear
correlation between $r/r_{\rm cool}$ and $t_{\rm cool}$.

Seven clusters have $r/r_{\rm cool}>1$, three have $r/r_{\rm cool}\sim1$
and six have $r/r_{\rm cool}<1$.  So in most (10 out of 16) of the clusters the
energy supplied by the bubbles is sufficient to offset the X-ray
cooling.  Of the six which have insufficient, four have young, active
bubbles.  A possible explanation is that these bubbles are still
growing and so have not reached their maximum size.  Therefore they currently do not contain sufficient energy to offset the cooling,
but may do at later stages in their evolution.  For further discussion
see Section \ref{bubble_size}.

The ghost bubbles are more difficult to explain as these bubbles would
be expected to be at their final size.  They are also those which
supply the least amount of energy into the ICM.  The cooling times at
the centres of these clusters are very short ($<1\gyr$) and so a
comparatively large amount of heating is required to offset the X-ray
cooling.  It is possible that we have caught the AGN in a period of relative
quiesence where the ICM is cooling and only small bubbles are
produced.  Once a reservoir of fuel for the AGN has built up from the
ICM then if it goes into a major outburst the bubbles may be such that
they offset the cooling.

\subsection{Bubble Sizes}\label{bubble_size}

So far we have assumed that the bubbles in the clusters are at their
``maximum'' size, however is this true?  Calculating the bubble radius
expected if its energy were offsetting all the X-ray cooling we show
that it is likely that our sample contains a large fraction of small bubbles.

If the jet supplying the
bubbles is constant, then there is expected to be a continuous cycle
of bubble growth with periodic detachment.  As the radius of the bubble depends on the
cube-root of the bubble age, most bubbles would be observed when they
are close to their
maximum sizes.  From \citet{Churazov00}, in the assumption that the jet
is of constant power and the bubbles are expanding subsonically,

\begin{eqnarray}
\frac{\gamma_1}{\gamma_1-1}P_{\rm th}V & = & {\mathcal C}Lt\nonumber\\
P_{\rm th} \frac{4}{3}\pi R_{\rm max}^3 & = & \frac{\gamma_1-1}{\gamma_1}{\mathcal C}Lt_{\rm max}\nonumber\\
R_{\rm max} & = & \sqrt{\frac{3{\mathcal C}L}{16\pi P_{\rm th} v_{\rm grav}}}.\label{Eq:rrmax}
\end{eqnarray}

\begin{figure}\centering

\includegraphics[width=1.0\columnwidth]{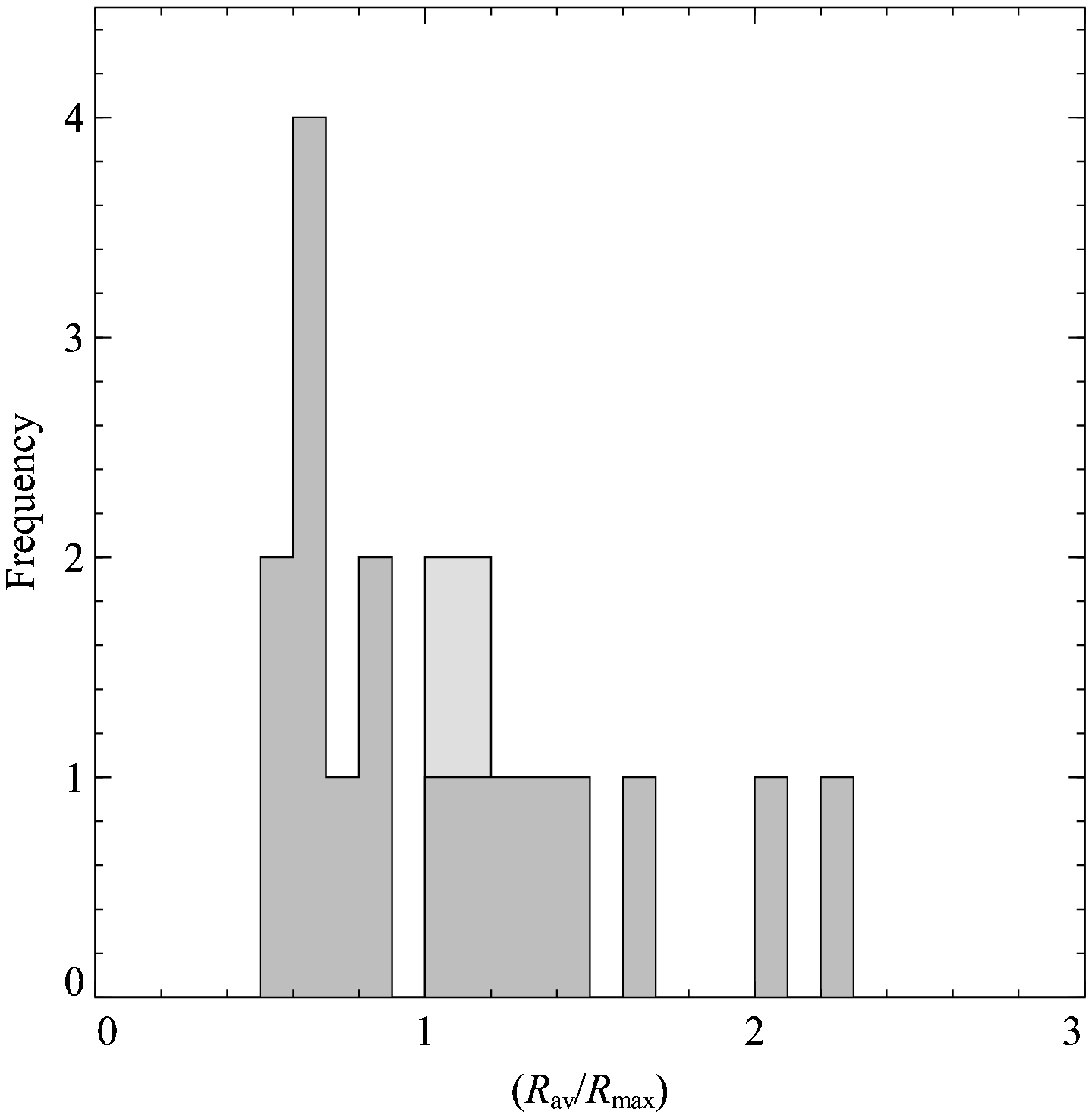}
\includegraphics[width=1.0\columnwidth]{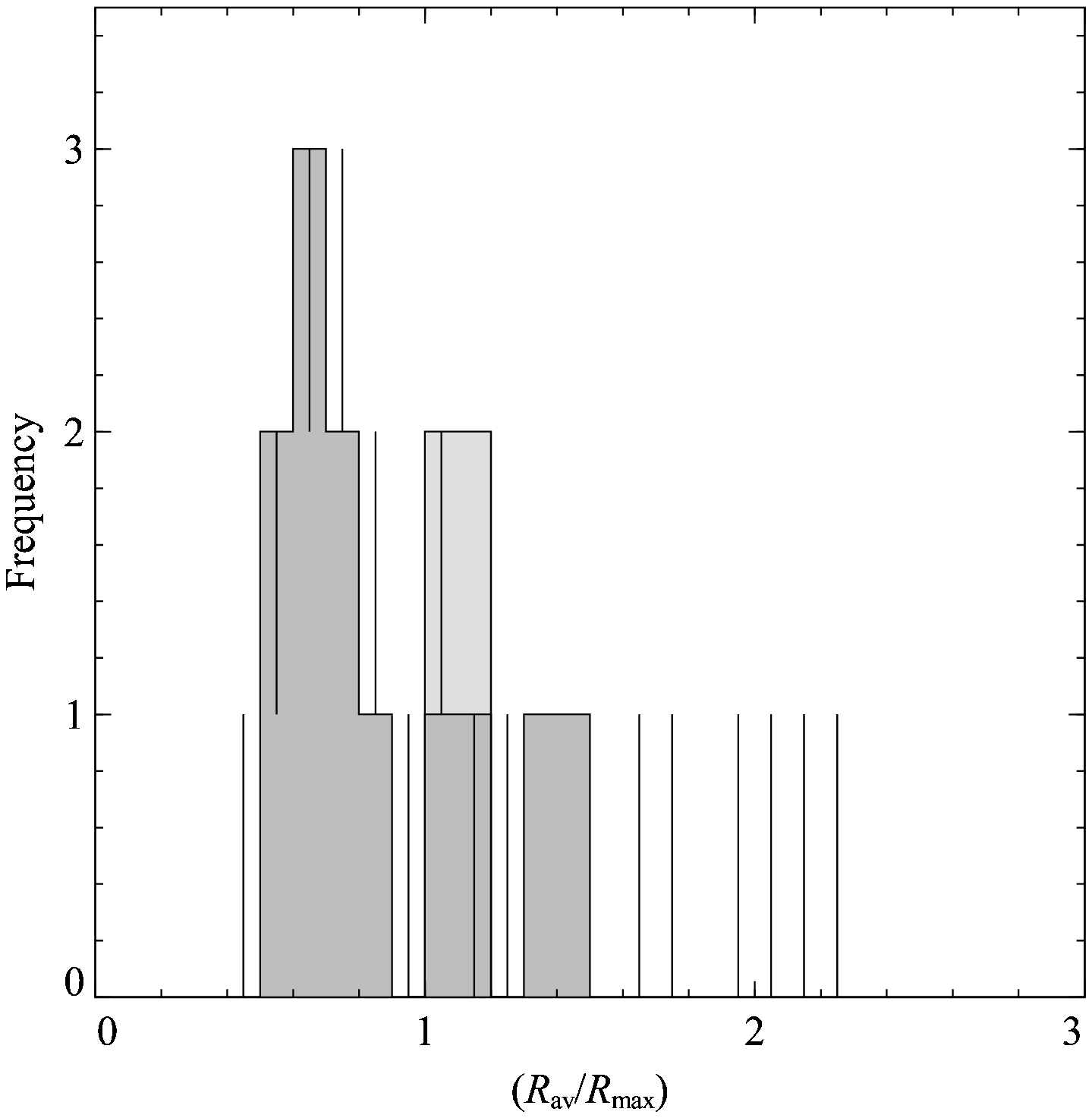}

\caption{\label{r_r_max} {\scshape Top:} The distribution of bubbles as a fraction of
  their maximum size $(R/R_{\rm max})$ assuming subsonic
  expansion. {\scshape Bottom:} The result of Monte Carlo simulations
  of the distribution in order to obtain errors (the vertical
  lines) on the distribution, and so a clearer indication of the
  true distribution.  The lighter grey bars are from M87.  The bubbles
from Cygnus A and one from A2052 would appear off to the right of this plot.}
\end{figure}

\noindent
We naively assume that the numerical constant, ${\mathcal C}=1$,
$\gamma_1=4/3$ and $t_{\rm max}=R_{\rm max}/v_{\rm grav}$.  Using the sound speed,
$v_{c_{\rm s}}$, for $v_{\rm grav}$ as this depends only on the
surrounding ICM and not the bubble size, and half the power required to balance cooling
for $L=L_{\rm cool}/2$ (as there are two bubbles per cluster), $R_{\rm max}$ is calculated.  This is the maximum size of the
bubble possible if all the energy of its creation is converted to $P_{\rm th}V$ work
on the ICM.  The distribution of $R_{\rm av}/R_{\rm max}$ is shown in
Fig. \ref{r_r_max}, where $R_{\rm av}=\sqrt[3]{R_{\rm l}R_{\rm w}^2}$,
the average radius of the observed bubble.  Monte Carlo simulations of the data
have been used to take into account the errors in the
distribution and provide an estimate on the values of the bins\footnote{The uncertainties on $R_{\rm av}/R_{\rm max}$
  result from the uncertainties in the input data, e.g. temperature
  profiles of the cluster.  To obtain an estimate on the uncertainties
on the values of the binned  $R_{\rm av}/R_{\rm max}$ distribution,
the results were binned many times ($10^4$ runs), assuming a Gaussian distribution
of uncertainties in $R_{\rm av}/R_{\rm max}$.  The inter-quartile range in the values
for each of the bins is what is shown by the error bars.}.  
All clusters with young bubbles (11 clusters, 22 bubbles) are included
in the analysis and Fig. \ref{r_r_max}.  As our analysis of M87 did not go far enough out in radius to obtain
$r_{\rm cool}$ we used the temperature, and electron density profiles in
\citet{Ghizzardi04} to calculate $R_{\rm av}/R_{\rm max}$.  The values
of $t_{\rm max}$ and $R_{\rm av}/R_{\rm max}$ are shown in Table \ref{rrmax_table}.

The distribution peaks around $\sim 0.7 R_{\rm av}/R_{\rm
  max}$ with a tail extending to larger $R_{\rm av}/R_{\rm max}$.  This
implies that a large number of the bubbles have not yet attained their maximum
size, and others are much larger than they ``should'' be.

The bubbles with a small $R_{\rm av}/R_{\rm max}$ are likely to be young bubbles
caught in the act of growing.  They are not particularly younger than
any other bubbles, but may be young for the cluster.  For example, the inner bubbles in A2199, which have
been analysed here, are very much smaller than the large scale radio emission, which also corresponds to
clear decrements in the X-ray emission from the ICM.  The large X-ray
holes have $R_{\rm av}\sim15\kpc$ giving an $R_{\rm av}/R_{\rm
  max}\sim5$, whereas the inner bubbles have $R_{\rm av}\sim1.6\kpc$ giving an $R_{\rm av}/R_{\rm
  max}\sim0.5$.  These bubbles are also those which do
not supply sufficient energy to the ICM, which is unsurprising as we
have used $L_{\rm cool}$ when calculating $R_{\rm max}$.  

As the bubbles with $R_{\rm av}/R_{\rm max}
\lesssim 1$ are likely to be young bubbles, it is possible that those bubbles with
$R_{\rm av}/R_{\rm max}\gtrsim 1$ may be those which have detached and
are expanding as they rise up through the ICM.  

Also,
some of the clusters with $R_{\rm av}/R_{\rm max}>1$ have morphologies
which are different to the younger ones. Hydra A may not have
``bubbles'' in the same way as the other clusters.  The cluster-scale
outburst reported in \citet{NulsenHydra04} shows that the radio source
has ploughed through the ICM rather than being confined by it; Cygnus
A and A4059 are similar -- there are X-ray decrements, but the synchrotron
plasma is not totally confined by the ICM.  It may be the case that
the bubbles in A2052 and Perseus are just in the process of detaching,
they still appear as young bubbles as the radio emission has not had
time to decay, but they are at the maximum possible size for the
cluster and are beginning to buoyantly rise.  The bubble dimensions we
have used for the Perseus cluster in this analysis are not those labelled as ``Inner
Bubbles'' in \citet{Precession}.  Using the radii for the Inner
bubbles in this analysis ($\sim4\kpc$) gives an $R_{\rm
  av}/R_{\rm max}\sim0.6$.

\begin{table*}
\caption{\label{rrmax_table}{\sc $r_{\rm av}/R_{\rm max}$}}
\begin{tabular}{lccccc}
\hline
\hline
Cluster&$R_{\rm av}$&$P$&$c_{\rm s}$&$t_{\rm max}$&$R_{\rm av}/R_{\rm  max}$\\
& \kpc&{$\rm\thinspace eV~cm^{-3}$}&$10^7$\cmps&$10^7~\yr$\\
\hline
	   
 A262	   &$  1.97 \pm  0.11 $&$  60.1 \pm  0.7 $&$  4.90 \pm 0.02 $&$    0.38\pm	0.01 $&$ 1.04 \pm 0.06 $ \\
   	   &$  2.11 \pm  0.12 $&$  60.1 \pm  0.7 $&$  4.90 \pm 0.02 $&$    0.38\pm	0.01 $&$ 1.11 \pm 0.07 $ \\
 A1795*	   &$  3.40 \pm  0.11 $&$ 224   \pm 11   $&$  9.62 \pm 0.21 $&$    0.51\pm	0.03 $&$ 0.68 \pm 0.04 $ \\
	   &$  3.35 \pm  0.11 $&$ 224   \pm 11   $&$  9.62 \pm 0.21 $&$    0.51\pm	0.03 $&$ 0.67 \pm 0.04 $ \\
 A2029*	   &$  3.23 \pm  0.34 $&$ 389   \pm 16   $&$ 10.9  \pm 0.2  $&$    0.42\pm	0.04 $&$ 0.70 \pm 0.09 $ \\
	   &$  3.79 \pm  0.30 $&$ 389   \pm 16   $&$ 10.9  \pm 0.2  $&$    0.42\pm	0.04 $&$ 0.81 \pm 0.10 $ \\
  A2052    &$  9.98 \pm  0.51 $&$  70.2 \pm  1.3 $&$  6.45 \pm 0.04 $&$    0.67\pm	0.01 $&$ 2.25 \pm 0.12 $ \\
	   &$ 15.0  \pm  0.8  $&$  70.3 \pm  1.4 $&$  6.45 \pm 0.04 $&$    0.67\pm	0.01 $&$ 3.38 \pm 0.18 $ \\
 A2199*	   &$  1.57 \pm  0.06 $&$ 148   \pm  5   $&$  7.46 \pm 0.10 $&$    0.40\pm	0.02 $&$ 0.52 \pm 0.04 $ \\
	   &$  1.68 \pm  0.06 $&$ 148   \pm  5   $&$  7.46 \pm 0.10 $&$    0.40\pm	0.02 $&$ 0.55 \pm 0.04 $ \\
 A4059	   &$  2.98 \pm  0.14 $&$  55.3 \pm  2.4 $&$  6.57 \pm 0.07 $&$    0.54\pm	0.03 $&$ 0.82 \pm 0.06 $ \\
	   &$  4.47 \pm  0.24 $&$  55.3 \pm  2.4 $&$  6.57 \pm 0.07 $&$    0.54\pm	0.03 $&$ 1.23 \pm 0.10 $ \\
Centaurus* &$  1.97 \pm  0.17 $&$ 108   \pm  1   $&$  4.75 \pm 0.01 $&$    0.58\pm	0.01 $&$ 0.70 \pm 0.06 $ \\
	   &$  2.25 \pm  0.16 $&$ 108   \pm  1   $&$  4.75 \pm 0.01 $&$    0.58\pm	0.01 $&$ 0.79 \pm 0.06 $ \\
Cygnus A   &$ 23.7  \pm  1.2  $&$ 448   \pm 44   $&$ 13.4  \pm 0.5  $&$    0.23\pm	0.02 $&$ 7.64 \pm 0.55 $ \\
	   &$ 25.1  \pm  1.2  $&$ 454   \pm 40   $&$ 13.4  \pm 0.5  $&$    0.23\pm	0.02 $&$ 8.12 \pm 0.57 $ \\
 Hydra	   &$  8.44 \pm  0.40 $&$ 204   \pm  6   $&$  8.53 \pm 0.12 $&$    0.48\pm	0.01 $&$ 2.01 \pm 0.10 $ \\
	   &$  6.85 \pm  0.38 $&$ 208   \pm 23   $&$  8.55 \pm 0.12 $&$    0.47\pm	0.03 $&$ 1.69 \pm 0.07 $ \\
Perseus    &$  8.15 \pm  0.00 $&$ 306   \pm  2   $&$  8.98 \pm 0.05 $&$    0.65\pm	0.01 $&$ 1.36 \pm 0.01 $ \\
	   &$  8.89 \pm  0.00 $&$ 306   \pm  2   $&$  8.98 \pm 0.05 $&$    0.65\pm	0.01 $&$ 1.49 \pm 0.01 $ \\   
\hline
M87	   &$  1.35 \pm  0.03 $&$ 272   \pm 22   $&$  6.54 \pm 0.20 $&$    0.19\pm	0.02 $& $ 1.04 \pm 0.07 $ \\
	   &$  1.51 \pm  0.03 $&$ 272   \pm 22   $&$  6.54 \pm 0.20 $&$    0.19\pm	0.02 $& $ 1.16 \pm 0.08 $ \\

\hline
\end{tabular}
\begin{quote} The average bubble radius $(R_{\rm av}$), ICM
  pressure, sound speed ($V_{\rm s}$) and $R_{\rm av}/R_{\rm  max}$
  for all the young bubbles in the sample.  The starred clusters are
  those in which the bubbles do not provide sufficient energy to
  offset the X-ray cooling.
\end{quote}
\end{table*}

If we assume that most bubbles that we observe are at their maximum
size, then we have overestimated $R_{\rm max}$.  To move the peak in
the distribution, such that $R_{\rm av}/R_{\rm max}\sim 1$, $R_{\rm max}$ has to reduce by a factor $\sim0.7$.
Which of the assumptions that we have used in the analysis could be
changed to reduce $R_{\rm max}$?

We have assumed that the timescale over which the bubbles heat the ICM
is the same as the sound speed timescale of the bubbles' formation.
Bubbles may grow in fits and starts \citep{Fabian_05_visc_cond}, the actual time taken to
expand the bubble may be shorter, as there are
periods where there is no expansion.  Depending on the exact nature of
the growth the true expansion time may be less than the one
calculated.  The energy dissipation time is also going to be different
to the bubble creation timescale (see Section \ref{timescales} for
more discussion on this topic).

We have assumed $\gamma_1=4/3$, but if it is possible for $\gamma_1\sim 1.1$ then the peak of the distribution
occurs at $R_{\rm av}/R_{\rm max}\sim 1$.  A smaller value of
$\gamma_1$ means that the same amount of energy creates a smaller
bubble than if $\gamma_1=4/3$.  Investigations into the
weak shock around the northern Perseus bubbles have shown it to be
isothermal \citep{ACF_Per_Mega_06}, which requires an effective $\gamma_2\sim1$ for the ICM.  Further investigations
are required to determine the true nature of $\gamma$ in the radio
bubbles and the ICM.

\citet{Sanders05} find non-thermal X-rays from the central regions of
the Perseus Cluster.  These most likely arise from magnetic fields and
cosmic rays in the ICM.  Using the more recent $900\ks$ observation
Sanders {\it et al.} (in prep) find that the non-thermal electron
pressure within a radius of $40\kpc$ is comparable to the total thermal pressure.  As a result the
total pressure against which the bubbles are expanding is about twice
that which has previously been estimated.  Amending Equation
\ref{Eq:rrmax} results in
\begin{equation}
R_{\rm max}'=\sqrt{\frac{3{\mathcal C}L}{16\pi (2 P_{\rm th}) v_{\rm grav}}}=\frac{R_{\rm max}}{\sqrt{2}}=0.7R_{\rm max}.
\end{equation}
If the non-thermal components are taken into account, then $R_{\rm
  max}$ reduces by the factor of $0.7$ required.  As the distribution
of $R_{\rm av}/R_{\rm max}$ peaks at around $0.7$ then it is likely
that a non-thermal component of the ICM exists and is important in The
innermost refions of all cool core clusters.  We note that the
possible presence of a non-thermal pressure component was suggested by
\citet{Voigt06} from determination of mass profiles in cool core clusters.

\subsection{Timescales}\label{timescales}

The timescales used to calculate the power output into the cluster by
the creation of bubbles have been those of the sound speed.  So
far no strong shocks have been observed around the rims of the
bubbles.  This means that they are currently expanding at less than the
sound speed.  We therefore have an upper limit on the bubble expansion
time, which has been used as the timescale over which the
bubbles deposit energy into the ICM.  

This is the bubble expansion timescale, however, rather than the
timescale over which the bubbles dissipate their energy in the ICM.
We have so far assumed that these two timescales are
equivalent.  But, as the bubbles can still be seen in the form of detached ghost bubbles
(e.g. Perseus \& Centaurus), bubbles last longer than this timescale.  As
such the energy is dissipated over a longer timescale than has been
assumed here.

If bubbles are continually produced, so the moment one detaches, or
shortly thereafter, another one starts to form, then the difference in
timescales is not a problem.  Although the energy dissipation occurs
over a longer timescale than the bubble creation, the net input of
energy into the central regions of the cluster occurs on the bubble
creation timescale.  However, if the duty cycle of the central engine
is short, so that bubbles are produced only rarely, then the energy
dissipation timescale is the more relevant.  Accurate estimates of the time-averaged energy
dissipation rate are vital to the study of the energy input into the
central regions of these clusters.  Our study above indicated that the
duty cycle is at least 70 per cent in clusters which require some form
of heating.  As there are a number of clusters which, although they do
not have clear bubbles, have complicated central morphologies or radio
emission, the duty cycle may be as high as 90 per cent.

In some clusters, Hydra A for example \citep{NulsenHydra04}, there
appears to have been a ``cluster scale outburst.''  In these cases the
outburst may have heated the central parts of the cluster sufficiently
so that it is not, currently, cooling very fast.  Therefore the
observed bubble size and the current cooling rate may not be as
closely coupled as we have assumed.  The heating effect of such large
scale requires further investigation, as in these cases the ICM does
not confine the bubble and the ICM pressure cannot be taken as uniform.

There exist radio sources where the morphology indicates that activity
has restarted in the centre after a time of quiescence \citep{Schoenmakers_00,Saripalli_02}.  Although for
some of the clusters presented here even a duty cycle of 100 per-cent
would be insufficient to provide the energy required to off-set the
X-ray cooling, for others a duty cycle of less than 100 per-cent would
suffice.  

For further improvements to the line of investigation we have taken,
the rate of the energy dissipation into the ICM, rather than the
energy injection, needs to be determined.

\subsection{Bubble Visibility}

The clusters with short cooling times are
also likely to be the ones which are X-ray bright as the gas, and
hence surface brightness, densities are high in the centre.  Therefore the
bubble-ICM contrast is also likely to be high.  This could cause a
selection effect for the percentage of clusters with short central
cooling times in which clear bubbles have been detected.  There may be
bubbles in clusters without short central cooling times, but as the
X-ray emission in the centre is not as bright, they have not been detected.

\section{Further Work}\label{furtherwork}

During this analysis we have assumed that the energy contained
within the bubble and which is available to the ICM is $4P_{\rm th}V$, assuming
that the bubbles are filled with a relativistic gas.  Apart
from energy lost to sound waves, which has been neglected in these
calculations \citep{Churazov02}, $4P_{\rm th}V$ is the maximum amount of energy
available to offset the X-ray cooling occurring in the cluster.  In some
cases this energy suffices to offset the cooling out to the cooling radius, in others it falls short.  

If, however, the energy were only $P_{\rm th}V$, then in many more clusters the
AGN would appear not to be able to offset the X-ray cooling.  As a
result, an important question that requires answering is whether there is $4P_{\rm th}V$ available in the bubble and whether
it can {\it all} be transferred to/dissipated in the ICM?

As discussed in Section \ref{timescales} the timescales used here to
obtain the bubble powers are those for the bubble age and not those
over which the bubbles would liberate their energy into the ICM.  They
also do not measure the time over which sound waves would dissipate.
These timescales also need to be accurately measured to subsequently
be able to accurately determine the heating rate of the AGN.

Recent work by \citet{ACF_Per_Mega_06}
investigating the weak shock surrounding the bubbles in the
Perseus cluster showed that the energy content of the post shock gas
was around $2P_{\rm th}V$.  The observation that the gas across the shock is
isothermal also raises the possibility that the bubbles could expand
faster than the sound speed.  Thermal conduction across the shock
would even out any temperature differences, so making the shock less detectable.  A
final issue is the energy in the compressed cosmic rays and magnetic
field present in the centres of clusters.  Sanders {\it et al.} in prep show that the thermal and non-thermal
pressures are comparable in the central regions of the Perseus cluster.  As a result, the pressure against which the
bubbles are expanding may be around double what has previously been
estimated.  The $R_{\rm av}/R_{\rm max}$ calculation (Section \ref{bubble_size}) implies that this non-thermal component
is important in other clusters.  As these inferences come
from a single cluster (albeit the X-ray brightest and best studied), further investigation is required to determine
whether these properties occur in all clusters and the effect that
they have on the assumptions used in AGN heating arguments.

\section{Conclusions}

We have investigated a sub-sample of the Brightest 55 cluster sample.
At least 36 per cent of clusters have a cool core, and of these at
least 70 per cent harbour clear bubbles.
First we analysed those clusters which have a short ($<3\gyr$) central
cooling time and a central temperature drop of a factor of two.  These
clusters are expected to have the most rapid cooling in the centre and
so require some form of heating.  At least 36 per cent of clusters have a cool core, and of these at
least 70 per cent harbour clear bubbles, implying a duty cycle also of
at least 70 per cent.  Up to 90 percent could have some form of bubbles.  The energy provided in the bubbles was compared to the X-ray
cooling within the cooling radius (for $r_{\rm cool}$ at $t_{\rm
  cool}=3\gyr$).   The mean
distance out to which the bubble power can offset the X-ray cooling is
$r_{\rm heat}/r_{\rm cool}\geq0.86\pm0.11$, with a large spread. In
most clusters (10 out of 16) the AGNs energy input to the central regions of the
cluster is sufficient to offset the X-ray cooling.  The bubble energy
in the remainder does not offset most of the X-ray cooling.  In some
cases this is the result of catching these bubbles at a very young age
(e.g. A2199).

Also analysed were those clusters in the B55 sample which have a
central radio source.  The fraction of clusters which have a central
radio source is at least 50 per cent. These were combined with the clusters from the
previous subset which did not have clear bubbles.  For clusters without clear bubbles we have used the mean value of
$r_{\rm heat}/r_{\rm cool}$ to estimate bubble sizes from the X-ray cooling.  In five clusters
(3C129.1, A2063, A2204, A3312 \& A3391) the radio
sources have bi-lobed morphologies, whose sizes are similar
from the expected sizes.  Only in two of these five clusters is the X-ray
observation of sufficient depth that any interaction between the
radio and ICM could be expected to be seen. It is not clear why no
clear bubbles are seen, especially in A2204.  As a result
bubbles may be more common than previously thought.

We investigated the ratio of the actual bubble size to that of their maximum size if they
were to offset all the X-ray cooling, and find the peak occurs at $R_{\rm
  av}/R_{\rm max}\sim 0.7$ with a tail extending to higher $R_{\rm
  av}/R_{\rm max}$.  Either we are biased to imaging young bubbles or
some assumption is incorrect.   A likely explanation is that a 
non-thermal component comparable to the thermal component is present
in the innermost regions of most clusters.

\section*{Acknowledgements}

We thank Steve Allen, Roderick Johnstone, Jeremy Sanders and James Graham for
interesting discussions during the course of this work.  ACF and RJHD
acknowledge support from The Royal Society and PPARC respectively.
All plots in this manuscript were created with {\it Veusz}.

\bibliographystyle{mnras} 
\bibliography{MN-06-0650-MJ-R1}

\section*{Appendix}

Here we present the temperature, density, pressure, entropy, cooling
time and cumulative heating profiles for the clusters in our sample
which harbour clear bubbles.  

\begin{figure}
\includegraphics[width=1.0\columnwidth]{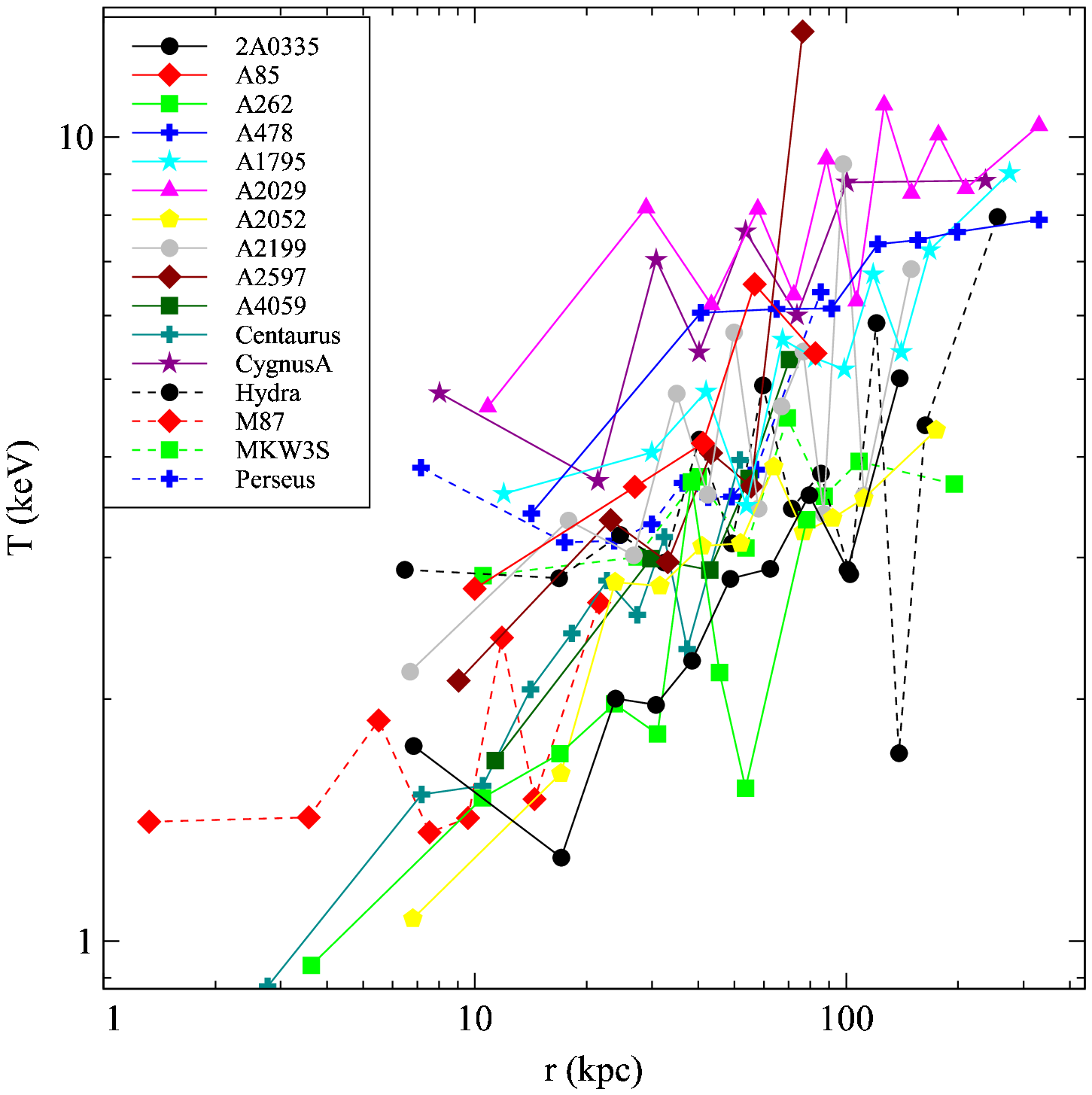}
\caption{\label{combined_temp}The temperature profiles of all the
  clusters in the sample which harbour clear bubbles.  Error bars have
been omitted for clarity.  The $1-\sigma$ errorbars are $\sim2-10$ per cent at $10\kpc$ and $\sim0.6-15$ per
cent at $100\kpc$.  The average powerlaw index is $0.32$ with a range of $0.11-0.59$.}
\end{figure}

As can be seen from the figures, the temperature profiles
(Fig. \ref{combined_temp}) do not lie on top of each other, but the
central temperature drops can be seen in these clusters.  The electron
density profiles are much more uniform and much smoother.  These
features carry into the derived profiles; the pressure's do not lie on
top of each other and fall as expected with radius. 

The entropy\footnote{We use $S=kTn_{\rm e}^{-2/3}$ as a proxy for the
  true entropy.} and cooling time profiles for the clusters are very
similar.  The scatter about the general trends are much less in these
profiles than in the others.  The overall shape of these profiles are
similar to those in \citet{Pratt_06, VoigtFabian04}.   

We have fitted simple powerlaws to all of the profiles of the clusters
individually.  The mean, maximum and minimum are given in the figure
captions.  In some cases the values for the outermost shell have been
ignored when this is obviously different from a global trend
\citep{Johnstone05}.  If a shell has not been used for a fit in the
denisty or temperature, it has not been used any of the fits in
subsequent derived profiles.

\begin{figure}
\includegraphics[width=1.0\columnwidth]{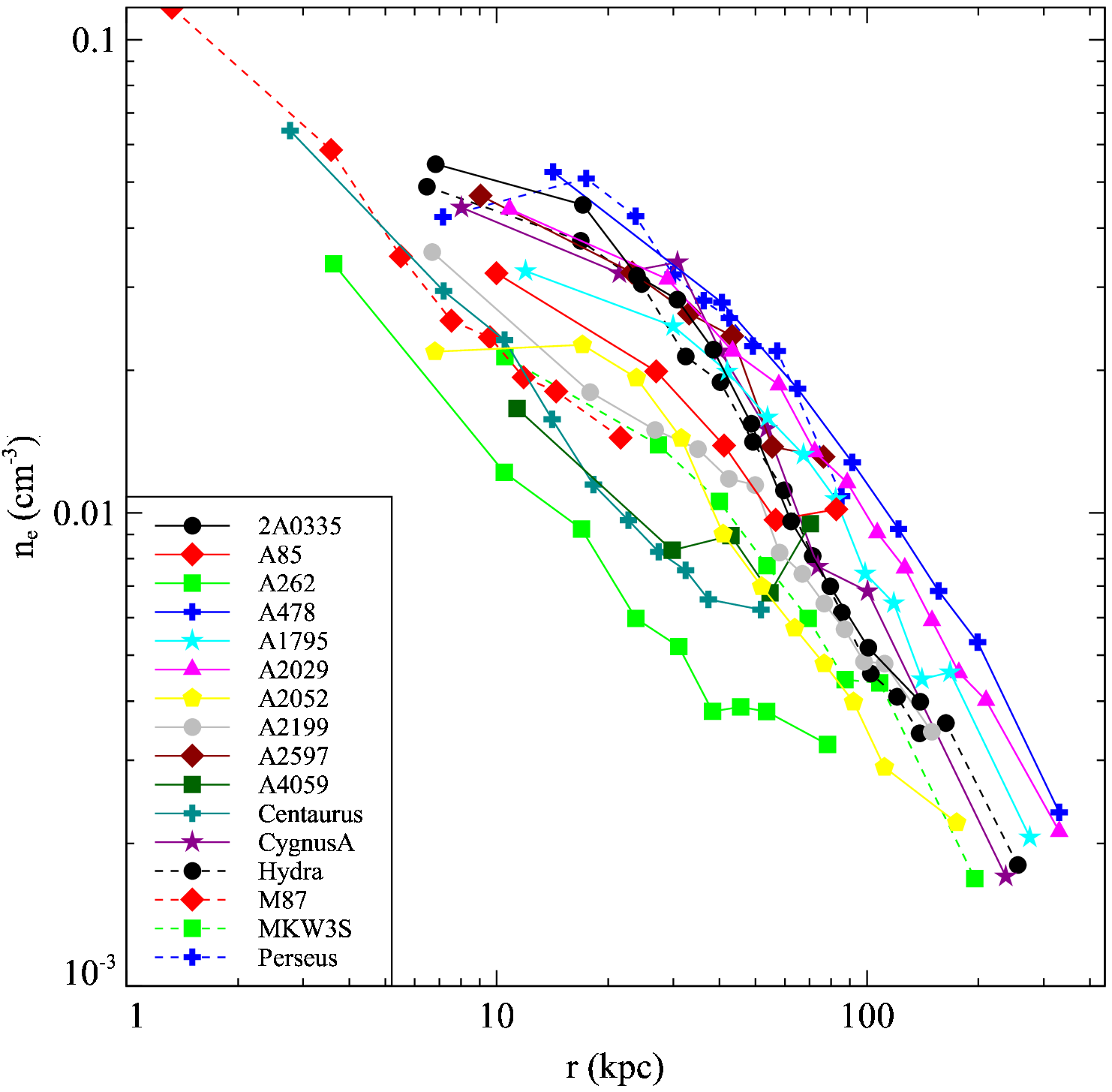}
\caption{\label{combined_ne}The electron density profiles of all the
  clusters in the sample which harbour clear bubbles.  Error bars have
been omitted for clarity.  The $1-\sigma$ errorbars are $\sim1-3$ per cent at $10\kpc$ and $\sim2$ per
cent at $100\kpc$.  The average powerlaw index is $-0.90$ with a range
of $-0.13$ to $-1.34$.}
\end{figure}
\begin{figure}
\includegraphics[width=1.0\columnwidth]{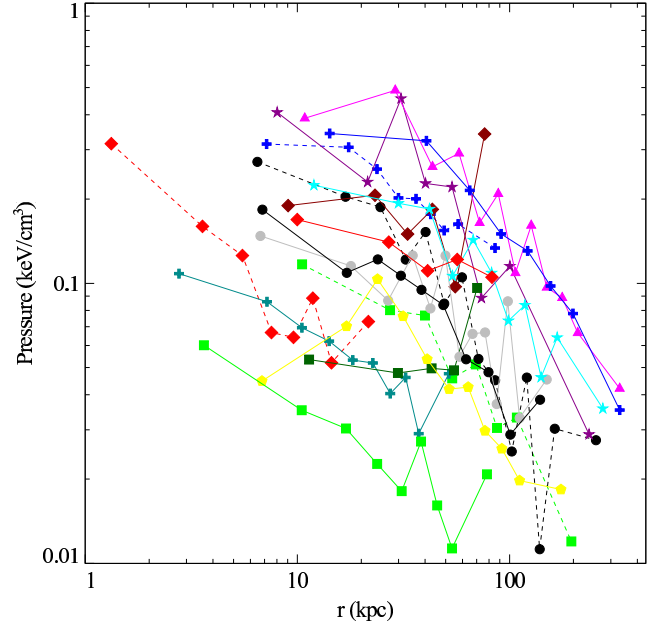}
\caption{\label{combined_press}The pressure profiles of all the
  clusters in the sample which harbour clear bubbles.  The legend is
  missing deliberately, but is the same as the other profiles.  The
  average powerlaw index is $-0.60$ with a range of $-0.09$ to $-1.13$.}
\end{figure}
\begin{figure}
\includegraphics[width=1.0\columnwidth]{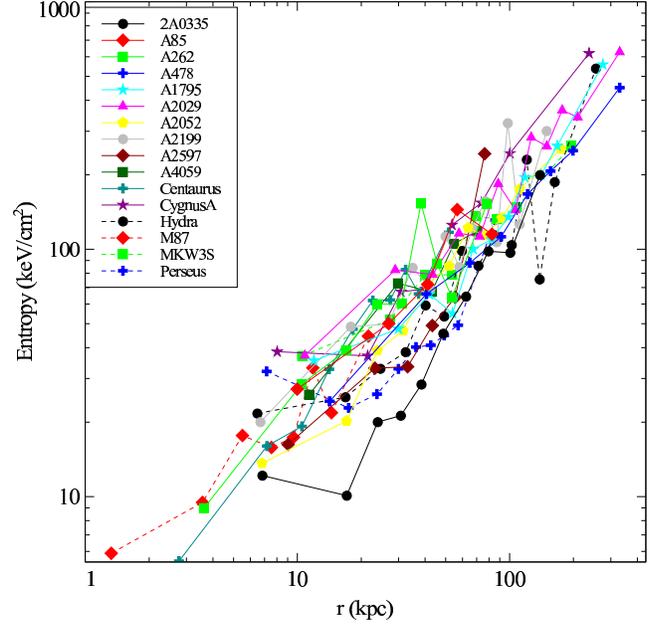}
\caption{\label{combined_entropy}The entropy profiles of all the
  clusters in the sample which harbour clear bubbles.  The average powerlaw index is $0.94$ with a range of $0.68-1.41$.}
\end{figure}
\begin{figure}
\includegraphics[width=1.0\columnwidth]{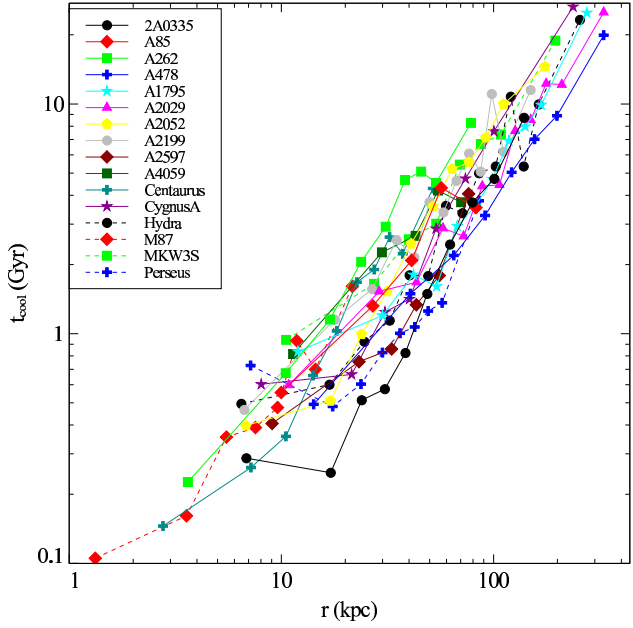}
\caption{\label{combined_tcool}The cooling time profiles of all the
  clusters in the sample which harbour clear bubbles.  The average powerlaw index is $1.19$ with a range of $0.83-1.76$.}
\end{figure}
\begin{figure}
\includegraphics[width=1.0\columnwidth]{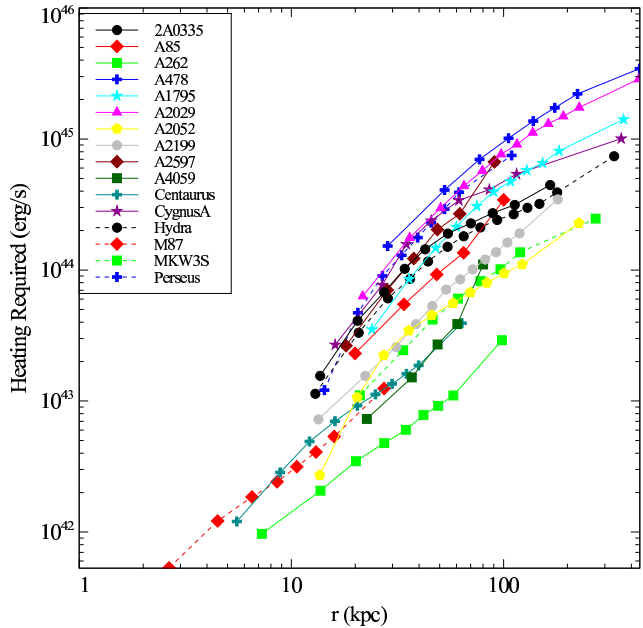}
\caption{\label{combined_heating}The profiles of all the
  clusters in the sample which harbour clear bubbles showing the
  cumulative amount of power required to offset the X-ray cooling.
  Note that the differential of these curves, the heating radio per
  kpc, is approximately flat (Fig. \ref{combined_heating_kpc}).  The
  average powerlaw index is $1.40$ with a range of $0.80-2.09$. }
\end{figure}
\begin{figure}
\includegraphics[width=1.0\columnwidth]{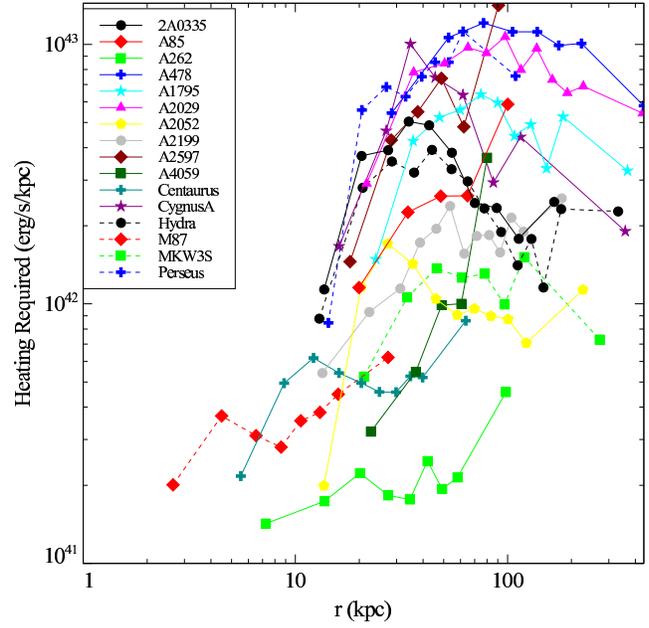}
\caption{\label{combined_heating_kpc}The profiles of all the
  clusters in the sample which harbour clear bubbles showing the
  amount of power required to offset the X-ray cooling per kpc.}
\end{figure}

\end{document}